\documentclass[apj, letterpaper]{emulateapj}  

\pdfoutput=1

\usepackage{amsmath}

\begin{document}

\slugcomment{\bf}
\slugcomment{ApJ, accepted and published}

\title{Doppler Signatures of the Atmospheric Circulation on Hot Jupiters}
\shorttitle{Atmospheric circulation of hot Jupiters}
\shortauthors{Showman et al.}

\author{Adam P. Showman\altaffilmark{1}, Jonathan
  J. Fortney\altaffilmark{2}, Nikole K. Lewis\altaffilmark{1,3}, and
  Megan Shabram\altaffilmark{4}}

\altaffiltext{1}{Department of Planetary Sciences and Lunar and Planetary
Laboratory, The University of Arizona, 1629 University Blvd., Tucson, AZ 85721 USA; 
showman@lpl.arizona.edu}
\altaffiltext{2}{Department of Astronomy and Astrophysics,
University of California, Santa Cruz, CA 95064, USA}
\altaffiltext{3}{Currently a Sagan Fellow at the Department of Earth, Atmospheric, and Planetary 
Sciences, Massachusetts Institute of Technology, Cambridge, MA 02139}
\altaffiltext{4}{Department of Astronomy, University of Florida,
211 Bryant Space Science Center, Gainesville, FL 32611-2055, USA}
\begin{abstract}
\label{Abstract}
The meteorology of hot Jupiters has been characterized primarily with
thermal measurements, but recent observations suggest the possibility
of directly detecting the winds by observing the Doppler shift of
spectral lines seen during transit. Motivated by these observations,
we show how Doppler measurements can place powerful constraints on the
meteorology. We show that the atmospheric circulation---and Doppler
signature---of hot Jupiters splits into two regimes. Under weak
stellar insolation, the day-night thermal forcing generates fast zonal
jet streams from the interaction of atmospheric waves with the mean
flow. In this regime, air along the terminator (as seen during
transit) flows toward Earth in some regions and away from Earth in
others, leading to a Doppler signature exhibiting superposed blue- and
redshifted components. Under intense stellar insolation, however, the
strong thermal forcing damps these planetary-scale waves, inhibiting
their ability to generate jets. Strong frictional drag likewise damps
these waves and inhibits jet formation. As a result, this second
regime exhibits a circulation dominated by high-altitude, day-to-night
airflow, leading to a predominantly blueshifted Doppler signature
during transit. We present state-of-the-art circulation models
including nongray radiative transfer to quantify this regime shift and
the resulting Doppler signatures; these models suggest that cool
planets like GJ~436b lie in the first regime, HD~189733b is
transitional, while planets hotter than HD~209458b lie in the second
regime. Moreover, we show how the amplitude of the Doppler shifts
constrains the strength of frictional drag in the upper atmospheres of
hot Jupiters. If due to winds, the $\sim$$2\rm\,km\,s^{-1}$ blueshift
inferred on HD 209458b may require drag time constants as short as
$10^4$--$10^6$ seconds, possibly the result of Lorentz-force braking
on this planet's hot dayside.

\end{abstract}



\section{Introduction}
\label{Introduction}

To date, the exotic meteorology of hot Jupiters has been characterized
primarily with thermal emission observations, particularly infrared
light curves \citep[e.g.][]{knutson-etal-2007b, 
  knutson-etal-2009a, knutson-etal-2012, cowan-etal-2007,
crossfield-etal-2010} and secondary eclipse
measurements \citep[e.g.,][]{charbonneau-etal-2005,
  deming-etal-2005a}.  Together, these observations place important
constraints on the vertical temperature profiles, day-night
temperature differences, and magnitude of day-night heat transport due
to the atmospheric circulation.  Moreover, in the case of HD 189733b
and Ups And b, infrared lightcurves indicate an eastward displacement
of the hottest region from the substellar longitude
\citep{knutson-etal-2007b, knutson-etal-2009a, crossfield-etal-2010}.
This feature is a common outcome of atmospheric circulation models,
which generally exhibit fast eastward windflow at the equator that
displaces the thermal maxima to the east \citep{showman-guillot-2002,
  cooper-showman-2005, showman-etal-2008a, showman-etal-2009,
  dobbs-dixon-lin-2008, dobbs-dixon-etal-2010, menou-rauscher-2009,
  menou-rauscher-2010, rauscher-menou-2010, rauscher-menou-2012b,
   burrows-etal-2010, thrastarson-cho-2010, lewis-etal-2010,
  heng-etal-2011, heng-etal-2011b, showman-polvani-2011,
  perna-etal-2012}.  In this way, the light curves provide
information---albeit indirectly---on the atmospheric wind regime.

Recent developments, however, open the possibility of direct
observational measurement of the atmospheric winds on hot Jupiters.
\citet{snellen-etal-2010} presented high-resolution groundbased, 
2-$\mu$m spectra obtained during the transit of HD 209458b in front of 
its host star.  From an analysis of 56 spectral lines of carbon monoxide, 
they reported an overall blueshift of
$2\pm1\rm\,km\,s^{-1}$ relative to the expected planetary motion,
which they interpreted as a signature of atmospheric winds flowing from
dayside to nightside toward Earth along the planet's terminator.
In a similar vein, \citet{hedelt-etal-2011} presented transmission spectra of
Venus from its 2004 transit, in which they detected Doppler shifted spectral
lines in the upper atmosphere, again seemingly the result of atmospheric winds.
These observations pave the way for an entirely new approach to
characterizing hot Jupiter meteorology.

The possibility of characterizing hot Jupiter meteorology via Doppler
provides a strong motivation for determining the types of Doppler
signatures generated by the atmospheric circulation.
\citet{seager-sasselov-2000} first mentioned the possible influence
of exoplanet winds on their transit spectra, and \citet{brown-2001}
considered the effect in more detail.  More recently,
\citet{kempton-rauscher-2012} took a detailed look at the ability of the
atmospheric circulation to affect the transmission spectrum.  Here, we
continue this line of inquire to show how Doppler measurements can
place powerful constraints on the meteorology of hot Jupiters.  We show that the
atmospheric circulation of hot Jupiters splits into two regimes---one
with strong zonal jets and superposed eddies, and the other comprising
predominant day-to-night flow at high-altitudes, with weaker
jets---which exhibit distinct Doppler signatures.

In Section~\ref{theory}, we present theoretical considerations 
demonstrating why two regimes should occur and the conditions for transition 
between them.  In Section~\ref{sw}, we test these ideas with an
idealized dynamical model.  Section~\ref{3D} presents
state-of-the-art three-dimensional dynamical models of three
planets---GJ 436b, HD 189733b, and HD 209458b---that bracket a wide
range of stellar irradiation and plausibly
span the transition from jet to eddy-dominated\footnote{Eddies refer
to the deviation of the winds from their zonal average.} at the low pressures
sensed by Doppler measurements.  Section~\ref{observables} presents
the expected Doppler signatures from these models, and Section~\ref{discussion}
concludes.

\section{Two regimes of atmospheric circulation: Theory}

\label{theory}

\begin{figure}
\includegraphics[scale=0.31, angle=0]{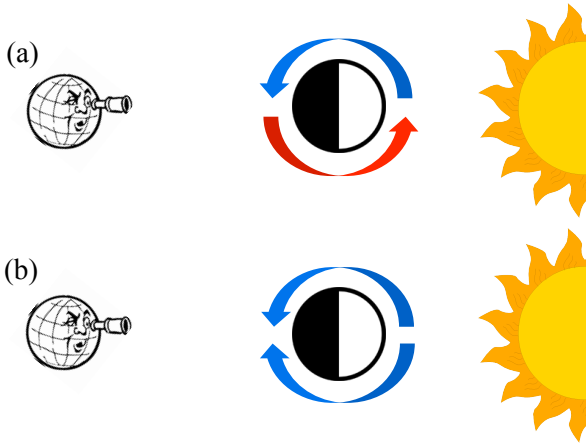}
\caption{Schematic implications of atmospheric circulation for
Doppler measurements of a hot Jupiter observed in transit.  Planet's
host star is depicted at right, planet (viewed looking down over
north pole) is at the center, and Earth is at the left. (a):  In
the presence of zonal jets, air flows along latitude circles (colored
arrows), leading to airflow toward Earth along one terminator (blue
arrow) and away from Earth along the other (red arrow).  A Doppler
signature that is broadened, or in extreme cases may be bimodally
split into blue- and redshifted components, results.  (b): When zonal
jets are damped, air flows primarily from day to night at low pressure,
leading to airflow toward Earth along both terminators (blue arrows).
A primarily blueshifted Doppler signature results.}
\label{cartoon}
\end{figure}

We expect the Doppler signature of the atmospheric circulation
on hot Jupiters to fall into two regimes, illustrated in 
Figure~\ref{cartoon}.

\subsection{Jet-dominated regime}
\label{jets-regime}

On rotating planets, the interaction of atmospheric turbulence with
the anisotropy introduced by the meridional gradient of the 
Coriolis parameter (known as the 
$\beta$ effect) leads to the emergence of zonal jets, which often
dominate the circulation (e.g., \citet{rhines-1975, rhines-1994},
\citet{williams-1978, williams-1979}, 
\citet{vallis-maltrud-1993}, \citet{cho-polvani-1996a} 
\citet{dritschel-mcintyre-2008}; for recent
reviews in the planetary context, see \citet{vasavada-showman-2005}
and \citet{showman-etal-2010}).   When radiative forcing and friction are weak,
the heating of air parcels as they cross the dayside or nightside will
be too small to induce significant day-night temperature variations;
the dominant driver of the flow will then be the meridional (latitudinal)
gradient in the
zonal-mean radiative heating.  Such a flow will exhibit significant
zonal symmetry in temperature and winds with the primary horizontal
temperature variations occurring between the equator and the poles.
For this regime to occur, the radiative timescale must be
significantly longer than the timescale for air parcels to cross a
hemisphere; the rotation rate must also be sufficiently fast, and
the friction sufficiently weak.  The speed and number of the zonal jets
will depend on a zonal momentum balance between Coriolis accelerations
acting on the mean-meridional circulation and eddy accelerations
resulting from baroclinic and/or barotropic instabilities, if any.  

When the radiative forcing is sufficiently strong, as expected for
typical hot Jupiters, large day-night heating contrasts will occur.
As shown by \citet{showman-polvani-2011}, such heating contrasts
induce standing, planetary-scale Rossby and Kelvin waves.  For typical
hot Jupiter parameters, these waves cause an equatorward flux of eddy
angular momentum that drives a superrotating (eastward) jet at the
equator \citep{showman-polvani-2011}.  This provides a theoretical
explanation for the near-ubiquitous emergence of eastward equatorial
jets in atmospheric circulation models of hot Jupiters.

In these jet-dominated regimes\footnote{The
interaction of eddies with the mean flow is generally responsible
for driving zonal jets, so eddies are almost never negligible to
the dynamics, even when zonal jets are strong.  
Here, by ``jet dominated'' we do not mean that eddies are unimportant
but rather simply that the resulting
jets have velocity amplitudes that significantly exceed the amplitude
of the eddies.}  (Fig.~\ref{cartoon}a), air along the
terminator---as seen during transit---flows toward Earth in some
regions and away from Earth in others.  This leads to a Doppler
signature where spectral lines are broadened, with minimal overall
shift in the central wavelength.  In extreme cases the Doppler signature
may be split into distinct, superposed blue- and redshifted
velocity peaks.

\subsection{Suppression of jets by damping}

The presence of sufficiently strong radiative or frictional damping
can suppress the formation of zonal jets, leading to a circulation
that at high altitudes is dominated by day-to-night flow rather than
jets that are quasi-symmetric in longitude (Fig.~\ref{cartoon}b).
Here, we demonstrate the conditions under which the mechanisms of
\citet{showman-polvani-2011} are suppressed.

\citet{showman-polvani-2011} identified two specific mechanisms
for the emergence of equatorial superrotation in models of
synchronously rotating hot Jupiters.  We consider each in turn.

\subsubsection{Differential zonal wave propagation}

As described above, the day-night thermal 
forcing on a highly irradiated, synchronously rotating planet 
generates standing, planetary-scale Rossby and Kelvin waves.
The Kelvin waves straddle
the equator while the Rossby waves exhibit pressure perturbations
peaking in the midlatitudes for typical hot Jupiter parameters.
The (group) propagation of Kelvin waves is to the east while that of
long Rossby waves is to the west; this differential zonal propagation
induces an eastward phase shift of the standing wave pattern near the equator
and a westward phase shift at high latitudes.  The result is a pattern
of eddy velocities (northwest-southeast in the northern
hemisphere and southwest-northeast in the southern hemisphere) that
causes an equatorward flux of eddy angular momentum.

If the radiative or frictional timescales are significantly
shorter than the time required for Kelvin and Rossby waves to
propagate over a planetary radius, the waves are damped,
inhibiting their zonal propagation and preventing the latitude-dependent
phase shift necessary for the meridional angular momentum fluxes.  
Therefore, this mechanism for generating zonal jets is suppressed when 
radiative or frictional damping timescales are sufficiently short.  The 
Kelvin-wave dispersion relation in the primitive equations\footnote{The
primitive equations are the standard equations for large-scale 
atmospheric flows in stably stratified atmospheres.  They are
a simplification of the Navier-Stokes equations wherein the vertical 
momentum equation is replaced with local hydrostatic balance, and are
valid when $N^2 \gg \Omega^2$ (where $\Omega$ is the planetary rotation
rate) and horizontal length scales greatly
exceed the vertical length scales.  These conditions are generally
satisified for the large-scale flow in planetary atmospheres, including
that on hot Jupiters.  See \citet{showman-etal-2010} or 
\citet[][Chapter 2]{vallis-2006} for a more detailed discussion.} 
\begin{equation}
\omega = {Nk\over \left({m^2 + {1\over 4H^2}}\right)^{1/2}}
\label{kelvin-dispersion}
\end{equation}
where $\omega$ is wave frequency, $k>0$ and $m$ are zonal and
vertical wavenumbers, respectively, $N$ is the Brunt-Vaisala
frequency, and $H$ is the scale height.  The fastest propagation
speeds occur in the limit of long vertical wavelength
($m\to 0$), which yields $\omega = 2NHk$ and thus phase
and group propagation velocities of $2NH$.  The propagation
time across a hemisphere is thus roughly $a/NH$, where $a$ is 
the planetary radius.  We thus
expect this jet-driving mechanism to be inhibited when
\begin{equation}
\tau_{\rm rad}\ll {a\over NH} \quad{\rm or}\quad \tau_{\rm drag}\ll{a\over NH}.
\end{equation}
For typical hot-Jupiter parameters ($a=10^8\rm\,m$, $H\approx
400\rm\,km$, and $N\approx3\times10^{-3}\rm\,s^{-1}$ appropriate to a
vertically isothermal temperature profile for a gravity of
$10\rm\,m\,s^{-2}$ and specific heat at constant pressure of
$1.3\times10^4\rm\,J\,kg^{-1}\,K^{-1}$), we obtain
$a/NH\sim 10^5\sec$.  Thus, this mechanism should be inhibited
when the radiative or drag time scales are much shorter than 
$\sim$$10^5\rm\,s$.

\subsubsection{Multi-way force balance}

\begin{figure*}
\begin{minipage}[c]{1.\textwidth}
\includegraphics[scale=0.55, angle=0]{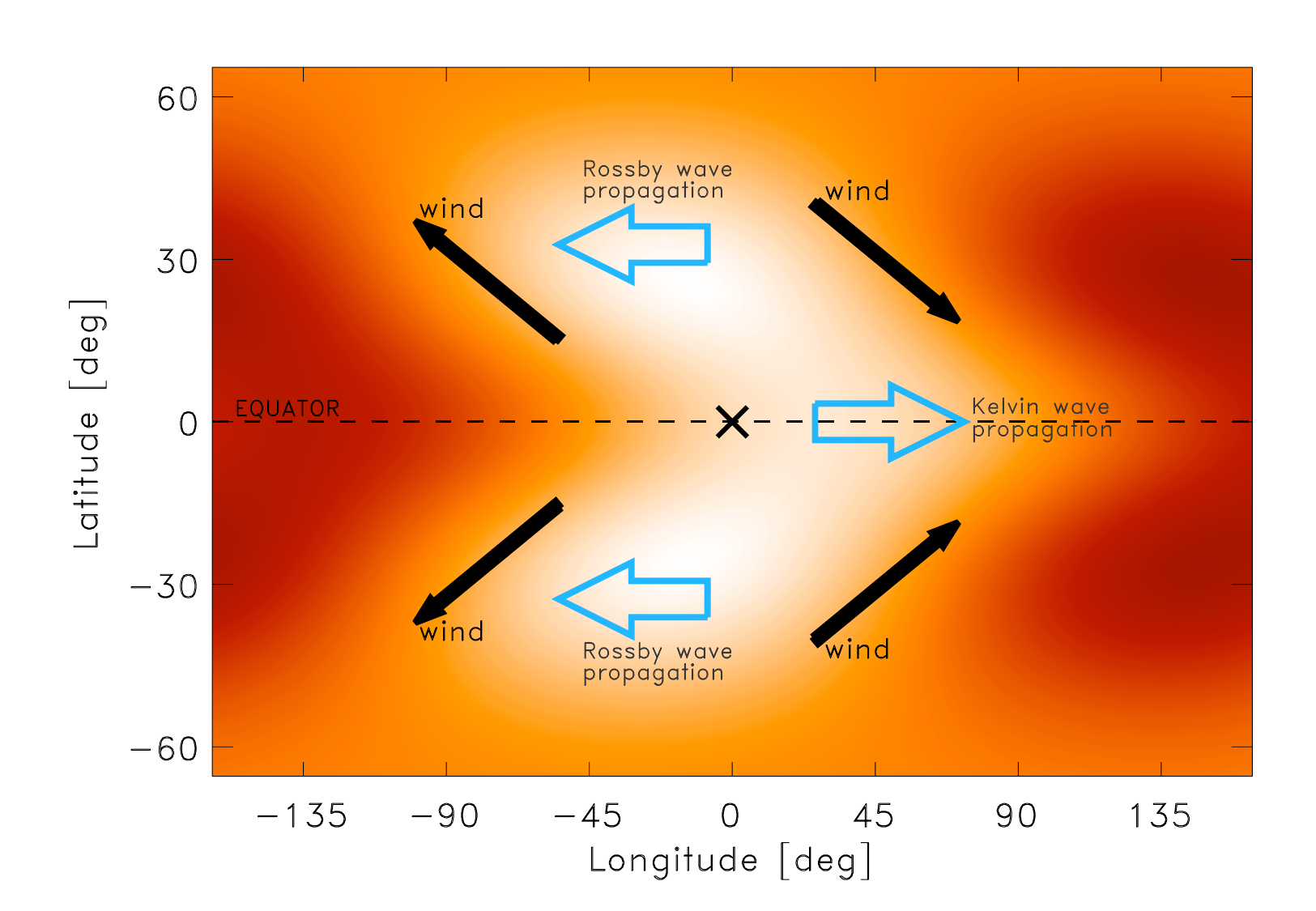}
\put(-240.,150.){\normalsize (a)}
\includegraphics[scale=0.55, angle=0]{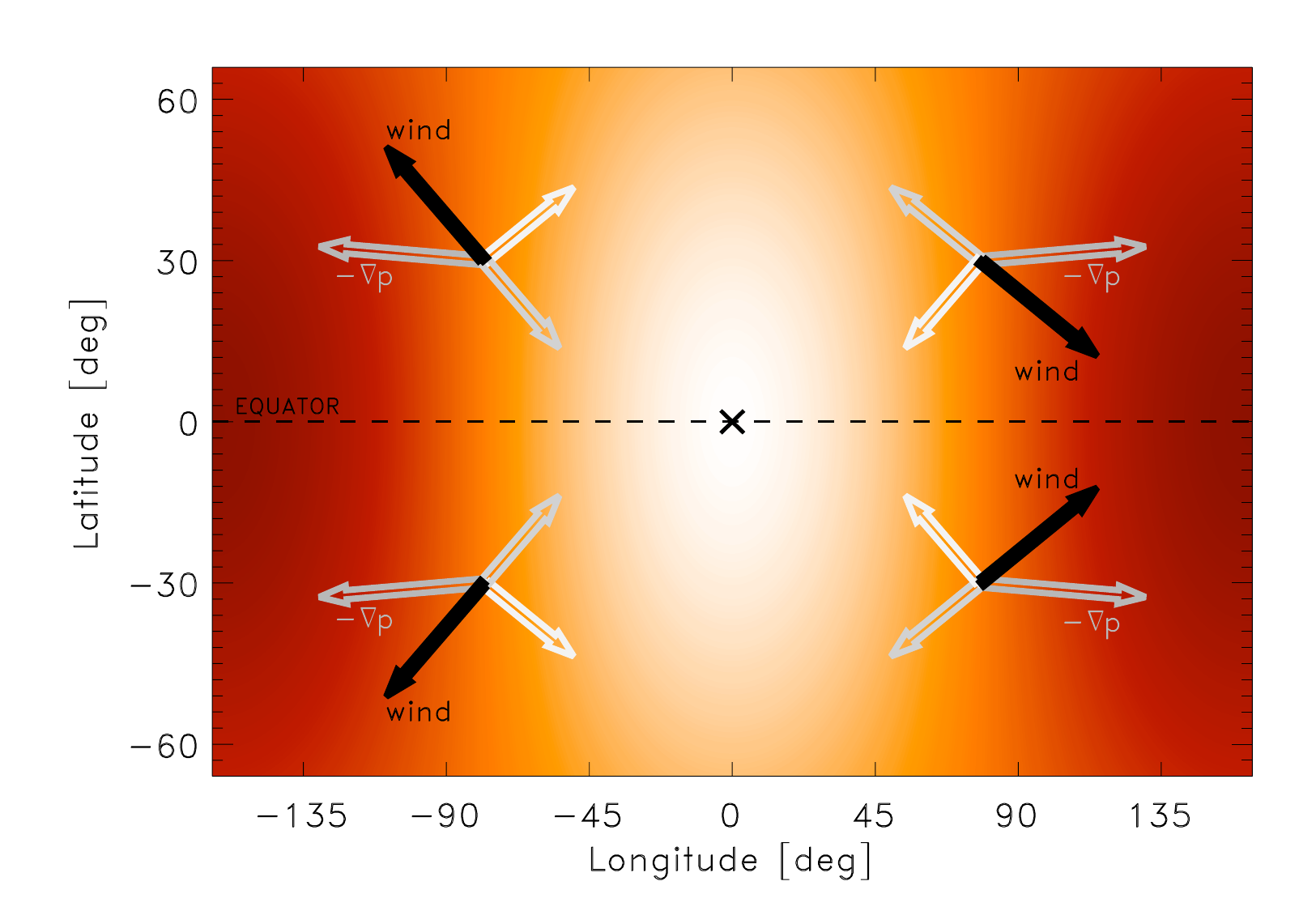}
\put(-240.,150.){\normalsize (b)}
{\linespread{1.0}
\caption{\small Schematic illustrating two mechanisms
    for driving equatorial superrotation on a hot Jupiter from
    \citet{showman-polvani-2011}.  (a) The day-night thermal forcing
    generates standing, planetary-scale Kelvin and Rossby waves.
    Differential zonal (east-west) propagation of these waves---Kelvin
    wave to the east and Rossby waves to the west---leads to an
    eastward displacement of the thermal structure at the equator and
    a westward displacement at midlatitudes and, in turn, eddy
    velocities that tilt northwest-southeast in the northern
    hemisphere and southwest-northeast in the southern hemisphere.
    This pattern leads to an equatorward flux of eddy momentum and the
    emergence of equatorial superrotation.  (b).  Even when radiative
    forcing is sufficiently strong to suppress the differential zonal
    thermal offsets just mentioned, a three-way force balance between
    pressure-gradient, Coriolis, and drag forces can lead to
    equatorward-eastward and poleward-westward velocity tilts, thereby
    driving equatorial superrotation.  Light regions indicate the
    dayside (with the substellar point marked by an $\times$), and
    darker regions indicate the nightside.  When the radiative time
    constant is short over a broad range of pressures, the
    pressure-gradient force points from day to night (long gray open
    arrows).  In the linear limit, the pressure-gradient force is
    balanced by the sum of drag (short gray open arrows) and Coriolis
    forces (short white open arrows).  The fact that the Coriolis
    force points to the right (left) of the wind vector in the
    northern (southern) hemisphere, and that drag typically points in
    the opposite direction of the wind itself, implies that drag and
    Coriolis forces will exhibit orientations qualitatively similar to
    those drawn in the figure when their amplitudes are comparable.
    This three-way force balance therefore implies that the wind
    vectors themselves exhibit an orientation which is rotated
    clockwise relative to $-\nabla p$ in the northern hemisphere and
    counterclockwise relative to $-\nabla p$ in the southern
    hemisphere.  The figure makes clear that, at low latitudes, these
    eddy wind-vector orientations correspond to northwest-southeast
    tilts in the northern hemisphere and southwest-northeast tilts in
    the southern hemisphere.  The result would be the equatorward
    transport of eddy angular momentum and the development of
    equatorial superrotation.\label{force-balance-schematic}}}
\end{minipage}
\end{figure*}

Even when the radiative timescale is extremely short and zonal propagation
of Rossby and Kelvin waves is inhibited, an eddy-velocity pattern that
promotes equatorial superrotation can occur under some conditions. As
pointed out by \citet{showman-polvani-2011} in the context of linear
solutions, a three-way horizontal force balance between pressure-gradient,
Coriolis, and frictional drag forces can lead to eddy velocities tilted
northwest-southeast in the northern hemisphere and southwest-northeast
in the southern hemisphere if the drag and Coriolis forces are comparable.
This occurs because drag generally points opposite to the wind direction,
whereas the  Coriolis force points to the right (left) of the wind in the 
northern (southern) hemisphere. 
When these two forces are comparable, balancing them with the 
pressure-gradient force requires
that the horizontal wind rotates clockwise of the day-night pressure-gradient
force in the northern hemisphere and counterclockwise of it in the
southern hemisphere (Figure~\ref{force-balance-schematic}).  
In the limit of short $\tau_{\rm rad}$ when
the horizontal pressure-gradient force points from day to night, these arguments
imply that, at low latitudes, the eddy velocities tilt 
northwest-southeast in the northern hemisphere and southwest-northeast
in the southern hemisphere \citep[Figure~\ref{force-balance-schematic}; see]
[Appendix D, for an analytic demonstration]{showman-polvani-2011}.

Even when frictional drag is too weak to play an important role
in the force balance, a similar three-way balance between pressure-gradient,
Coriolis, and advection forces can under appropriate conditions lead to
velocity tilts that promote equatorial superrotation.  As air flows
from day to night, the Coriolis force will deflect the trajectory
of the airflow to the right of the pressure-gradient force in the 
northern hemisphere and to the left of it in the southern hemisphere.
When the pressure-gradient force per mass, Coriolis force per mass, 
and advective acceleration are all comparable, as expected under
the Rossby number $Ro\sim1$ conditions typical of hot Jupiters,
then the deflection will be substantial.   In the limit of 
short $\tau_{\rm rad}$ when the horizontal pressure-gradient force 
points from day to night, these arguments again imply that the eddy 
velocities tilt northwest-southeast in the northern hemisphere and 
southwest-northeast in the southern hemisphere.

Now consider the effect of damping on this mechanism. To the degree
that the radiative timescale is short enough for temperatures to be
close to radiative equilibrium, {\it radiative} damping will not inhibit
this mechanism; however, strong {\it frictional} damping can prevent
it from occurring.  When the frictional force is much stronger than
the Coriolis and advective forces, the horizontal force balance is no
longer a multi-way force balance but rather becomes essentially a
two-way balance between the pressure-gradient force and drag.  In
this case, winds simply flow down the pressure gradient from day to night.
There is thus no overall tendency for prograde eddy-velocity tilts to
develop, so the jet-pumping Reynolds stress, and the jets themselves, are weak.

To quantify the amplitude of drag needed for this transition to occur,
consider a drag force per mass parameterized
by $-{\bf v}/\tau_{\rm drag}$, where ${\bf v}$ is horizontal velocity
and $\tau_{\rm drag}$ is the drag time constant.  The drag force
dominates over the Coriolis force when $\tau_{\rm drag}\ll f^{-1}$,
where $f=2\Omega\sin\phi$ is the Coriolis parameter, $\Omega$ is
the planetary rotation rate ($2\pi$ over the rotation period), and
$\phi$ is latitude.
Models of hot Jupiters predict flows whose dominant length scales
are global, in which case the
advective acceleration should scale as $U^2/a$, where $U$ is the
characteristic horizontal wind speed.  Drag will then
dominate over the advection force when $\tau_{\rm drag}
\ll(a/|\nabla\Phi|)^{1/2}$, where $|\nabla\Phi|$ is the characteristic
amplitude of the horizontal day-night pressure-gradient force on
isobars\footnote{The condition for dominance of drag over advection
can be motivated as follows.  When advection and drag are comparable, and both
together balance the pressure-gradient force, it
implies to order-of-magnitude that
\begin{equation}
{U^2\over a}\sim {U\over\tau_{\rm drag}}\sim |\nabla\Phi|.
\end{equation}
These two relations yield $\tau_{\rm drag}\sim a/U$ and
$U\sim\tau_{\rm drag}|\nabla\Phi|$, which together imply $\tau_{\rm
  drag}\sim(a/|\nabla\Phi|)^{1/2}$.  For drag time constants
significantly shorter than this value, the drag force exceeds the
advection force.}, given to order of magnitude by $|\nabla\Phi|\sim
R\Delta T_{\rm horiz}\Delta\ln p/a$, where $R$ is the specific gas
constant, $\Delta T_{\rm horiz}$ is the characteristic day-night
temperature difference, and $\Delta\ln p$ is the range of $\ln p$ over
which this temperature difference extends. For typical hot Jupiter
parameters, both conditions imply drag dominance for $\tau_{\rm
  drag}\ll 10^5\sec$.  When this condition is satisfied, the
horizontal force balance is between the pressure-gradient and drag
forces.  As mentioned above, the resulting circulation at low pressure
involves day-to-night flow with minimal zonal-mean eddy-momentum flux
convergences in the meridional direction and weak zonal jets.

\subsubsection{Direct damping of jets by friction}

Frictional drag can also directly damp the zonal jets.  A robust
understanding of how drag influences the equilibrated jet speed---and
hence a rigorous theoretical prediction of the amplitudes of drag
needed to damp the jet---requires a detailed theory for the full,
three-dimensional interactions of the global-scale planetary waves
with the background flow, which is currently lacking.  It is therefore
not possible at present to provide a robust theoretical estimate of
the amplitude of drag necessary to damp the zonal jets.  Still,
because the jets are fundamentally driven by global-scale waves that
result from the day-night heating gradients
\citep{showman-polvani-2011}, and because the radiative time constant
increases rapidly with depth, we expect that the magnitude of
zonal-mean acceleration of the zonal-mean zonal wind varies strongly
with depth.  These arguments heuristically suggest that the necessary
frictional damping times are less than a value ranging from $10^4\sec$
at low pressures of say  $\lesssim$0.1 bar to $10^6\sec$
or more at pressures of several bars, below the infrared photosphere.

\subsubsection{Recap}

When the jets and the waves that generate them are suppressed, 
the planet will tend to exhibit a large
day-night temperature difference at low pressure, resulting in a 
large horizontal pressure gradient force between day to night
that will drive a day-night flow (modified by the Coriolis effect)
at low pressure.  In this regime, air flows toward Earth
along most of the terminator, leading to a predominantly blueshifted Doppler
signature during transit.  Mass continuity requires the existence of
a return flow from night to day in the deep atmosphere (below the regions
sensed by Doppler transit measurements).  Because the
density at depth is much larger than that aloft, the velocities of
this return flow can be small.

\section{Test of the two regimes with an idealized model}
\label{sw}

We now demonstrate this transition from jet- to eddy-dominated
circulation regimes in an idealized dynamical
model.  As in \citet{showman-polvani-2011}, we consider a two-layer
model, with constant densities in each layer; the upper layer
represents the stratified, meteorologically active atmosphere and the
lower layer represents the denser, quiescent deep interior.  When the
lower layer is taken to be infinitely deep and the lower-layer winds
and pressure field are steady in time, the governing equations are the
shallow-water equations for the flow in the upper layer:
\begin{equation}
{d{\bf v}\over dt}+g\nabla h + f{\bf k}\times {\bf v} = {\bf R} 
- {{\bf v}\over{\tau_{\rm drag}}}
\label{momentum}
\end{equation}
\begin{equation}
{\partial h\over\partial t} + \nabla\cdot ({\bf v}h) = 
{h_{\rm eq}(\lambda,\phi) - h\over\tau_{\rm rad}}
\equiv Q
\label{continuity}
\end{equation}
where ${\bf v}(\lambda,\phi,t)$ is horizontal velocity,
$h(\lambda,\phi,t)$ is the upper layer thickness, $\lambda$ is
longitude, $t$ time, $g$ is the (reduced) gravity,\footnote{The
  reduced gravity is the gravity times the fractional density
  difference between the two layers.  For a hot Jupiter with a
  strongly stratified thermal profile, where entropy increases
  signficantly over a scale height, the reduced gravity is comparable
  to the actual gravity.} and $d/dt \equiv {\partial/\partial t} +
{\bf v}\cdot\nabla$ is the material (total) derivative.  The term
${\bf R}$ in Eq.~(\ref{momentum}) represents momentum advection
between the layers; it is $-{\bf v}Q/h$ in regions of heating ($Q>0$)
and zero in regions of cooling ($Q<0$).  See
\citet{showman-polvani-2011} for further discussion and interpretation
of the equations.

In the context of a three-dimensional (3D) atmosphere, the boundary
between the layers represents an atmospheric isentrope, and radiative
heating/cooling, which transports mass between layers, is therefore
represented as a mass source/sink, $Q$, in the upper-layer equations.
We parameterize this as a Newtonian cooling that relaxes the thickness
toward a radiative-equilibrium thickness, $h_{\rm eq}(\lambda,\phi)$,
over a prescribed radiative time constant $\tau_{\rm rad}$.  Here, we
set
\begin{equation}
h_{\rm eq}(\lambda,\phi) = \begin{cases}
  H  &{\rm on\; the\; nightside;}\\
  H + \Delta h_{\rm eq}\cos\lambda\cos\phi &{\rm on\; the\; dayside}
\end{cases}
\label{heq}
\end{equation}
where the substellar point is at
$(\lambda,\phi)=(0^{\circ},0^{\circ})$.  This expression incorporates
the fact that, on the nightside, the radiative equilibrium temperature
profile of a synchronously rotating hot Jupiter is constant
\citep[e.g.,][]{showman-etal-2008a}, whereas on the dayside the
radiative-equilibrium temperature increases from the terminator to the
substellar point.  An important property of Eq.~(\ref{heq}) is that
the {\it zonal-mean} radiative-equilibrium thickness, $\overline{h_{\rm
    eq}}$, is greater at the equator than the poles, reflecting the
fact that a planet with zero obliquity (whether tidally locked or not)
absorbs more sunlight at low latitudes than high latitudes.
Note that Eq.~(\ref{heq}) differs from the formulation of $h_{\rm eq}$
adopted by \citet{showman-polvani-2011}, where $h_{\rm eq}$ was set to 
$H + \Delta
h_{\rm eq}\cos\lambda\cos\phi$ across the entire planet (dayside and
nightside).

In addition to radiation, we include frictional drag parameterized with Rayleigh
friction, $-{\bf v}/\tau_{\rm drag}$, where $\tau_{\rm drag}$ is a specified
drag timescale.  The drag could result from vertical turbulent mixing
\citep{li-goodman-2010}, Lorentz-force braking \citep{perna-etal-2010}, 
or other processes.  

Our model formulation 
is identical to that described in \citet[][Section 3.2]{showman-polvani-2011}
in all ways except for the prescription of $h_{\rm eq}(\lambda,\phi)$.

Parameters are chosen to be appropriate for hot Jupiters.
We take $gH=4\times10^6\rm\,m^2\sec^{-2}$ and set $\Delta h_{\rm eq}/H=1$, 
implying that the radiative-equilibrium temperatures vary by order-unity 
from nightside to dayside.  We also take $\Omega=3.2\times10^{-5}\sec^{-1}$
and $a=8.2\times10^7\rm \,m$, implying a rotation period of 2.2 Earth days
and radius of 1.15 Jupiter radii, similar to the values for HD 189733b.
The radiative and frictional timescales are varied over a wide range
to characterize the dynamical regime.

We solved Equations~(\ref{momentum})--(\ref{continuity}) in full
spherical geometry using the Spectral Transform Shallow Water Model
(STSWM) of \citet{hack-jakob-1992}.  The equations are integrated
using a spectral truncation of T170, corresponding to a resolution
of $0.7^{\circ}$ in longitude and latitude (i.e., a global grid
of $512\times256$ in longitude and latitude).  All models were
integrated until a steady state is reached.

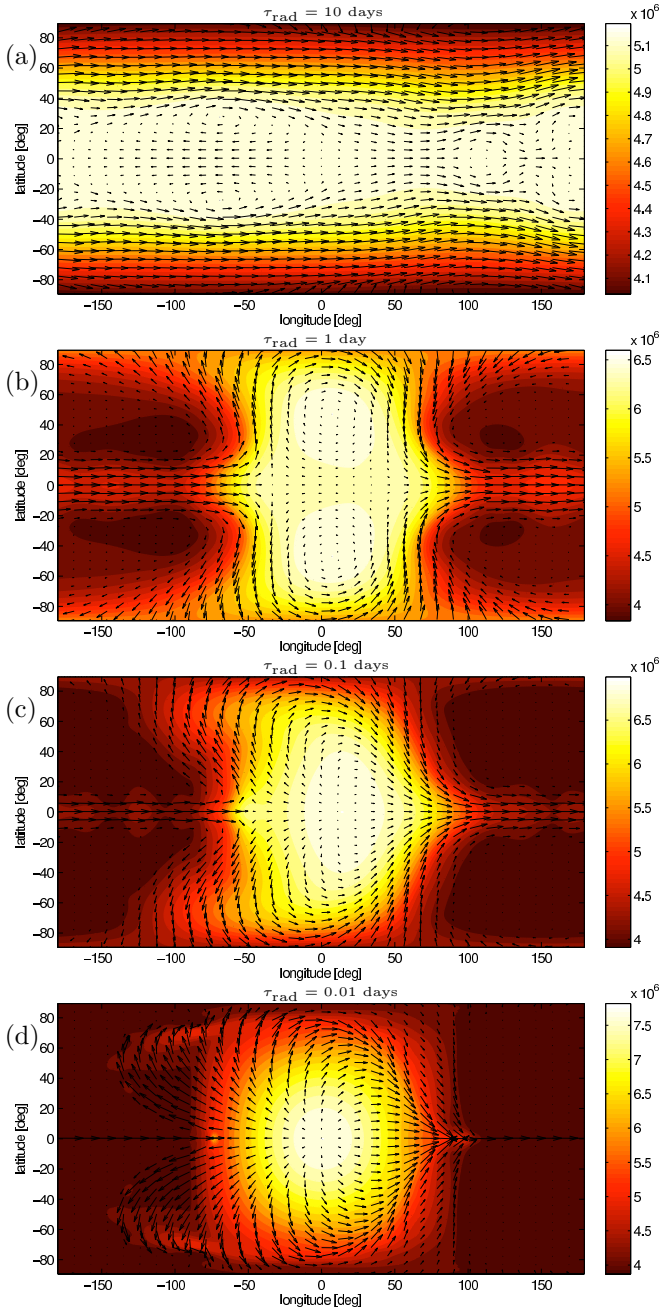
\begin{figure}
\begin{minipage}[c]{0.5\textwidth}
\includegraphics[scale=0.47, angle=0]{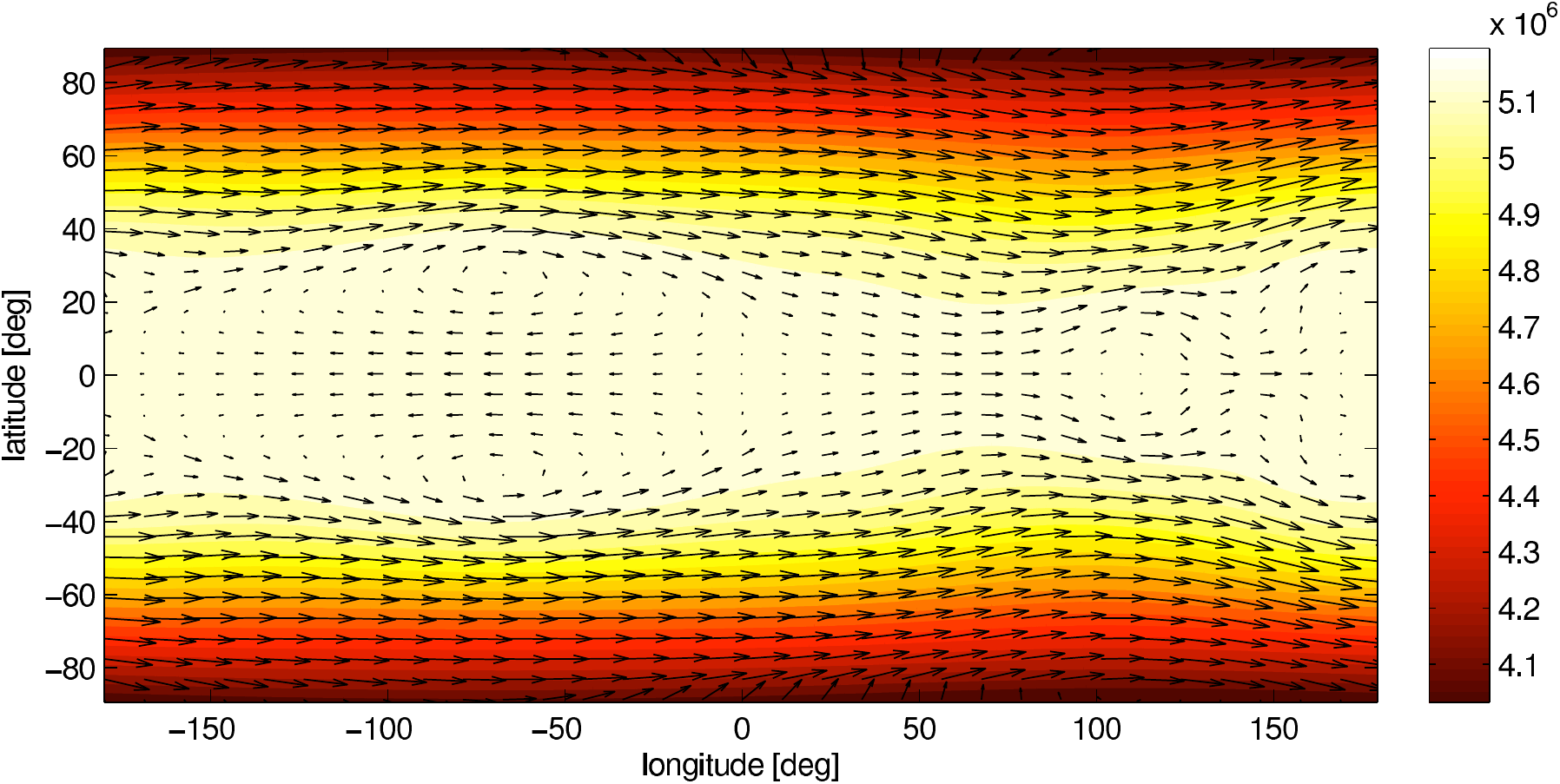}
\put(-251.,100.){\normalsize (a)}
\put(-150.,118.){\tiny $\tau_{\rm rad}=10$ days}

\includegraphics[scale=0.47, angle=0]{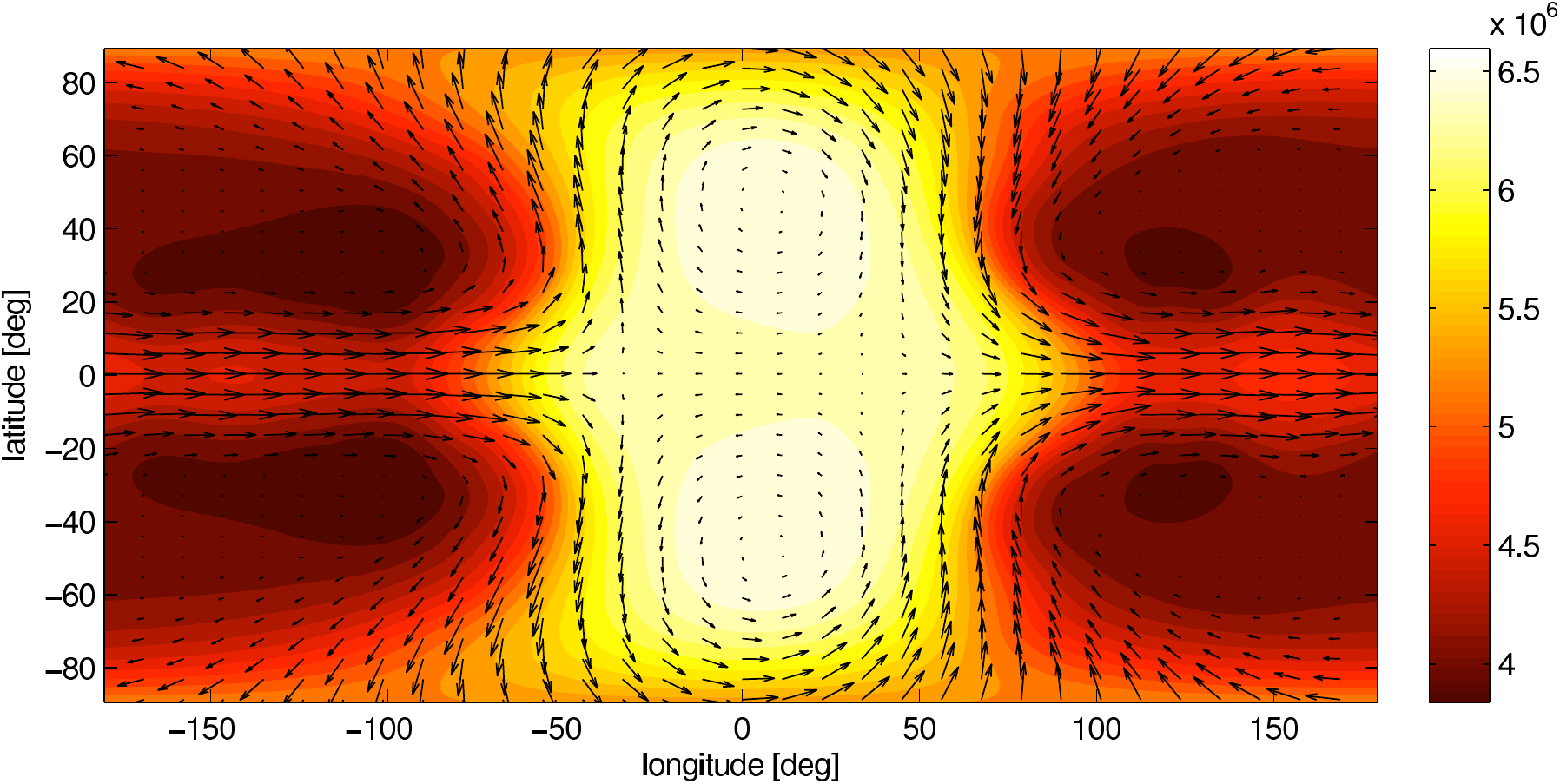}
\put(-251.,100.){\normalsize (b)}
\put(-150.,118.){\tiny $\tau_{\rm rad}=1$ day}

\includegraphics[scale=0.47, angle=0]{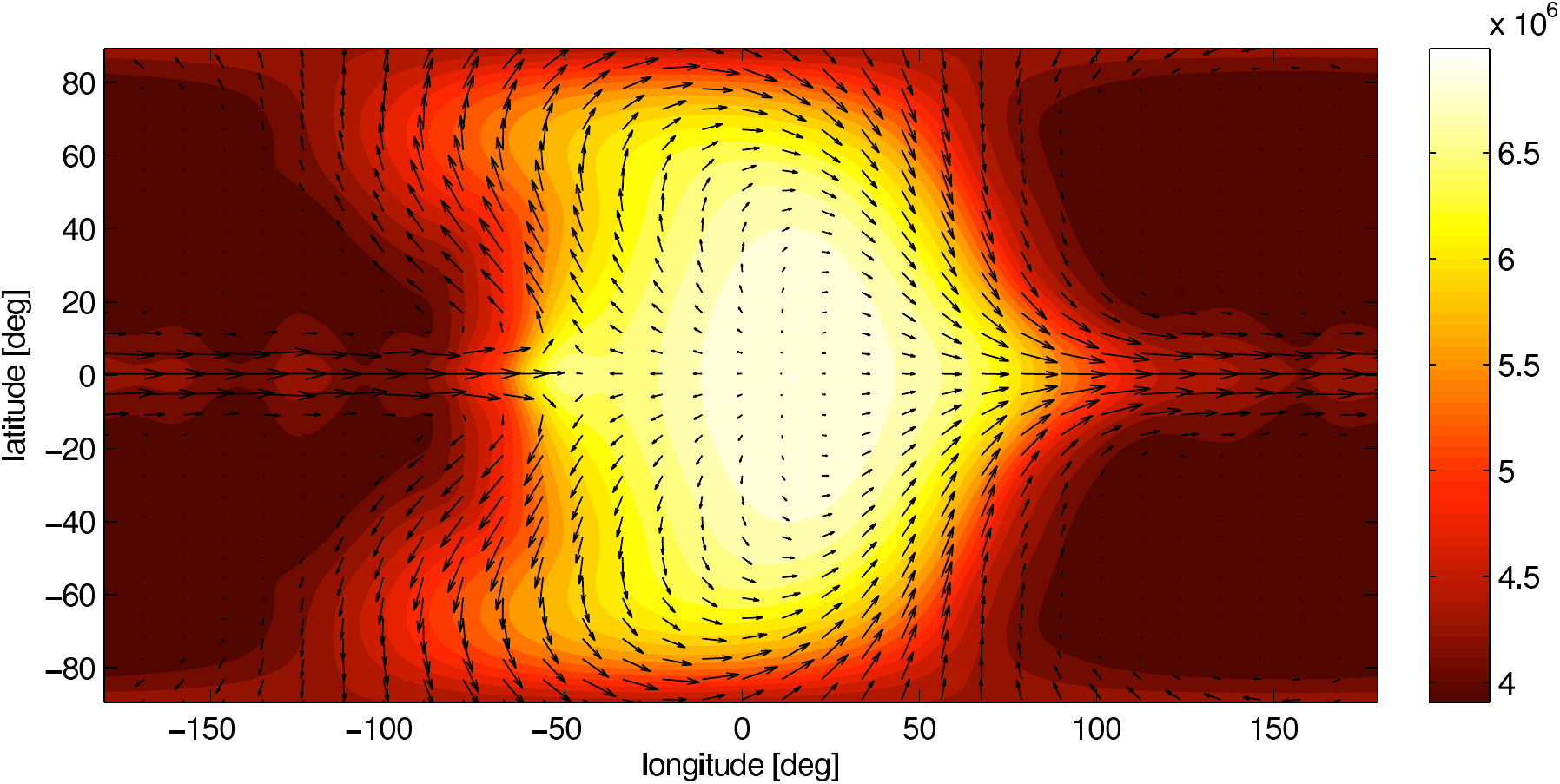}
\put(-251.,100.){\normalsize (c)}
\put(-150.,118.){\tiny $\tau_{\rm rad}=0.1$ days}

\includegraphics[scale=0.47, angle=0]{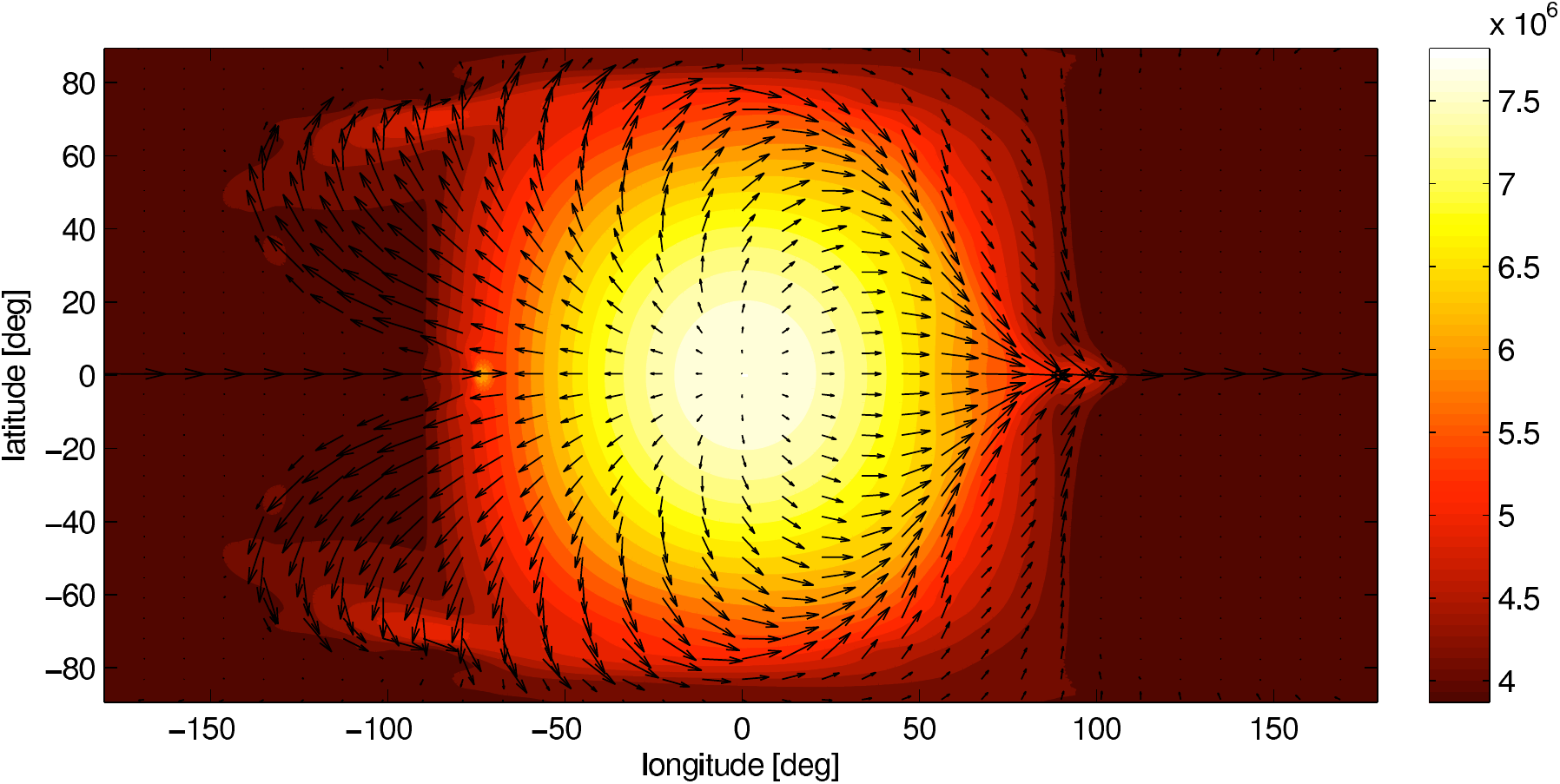}
\put(-251.,100.){\normalsize (d)}
\put(-150.,118.){\tiny $\tau_{\rm rad}=0.01$ days}

\caption{Geopotential $gh$ (orange scale, units $\rm m^2\,s^{-2}$) and winds (arrows) for
the equilibrated (steady-state) solutions to the shallow-water equations
(Equations~\ref{momentum}--\ref{continuity}) in full spherical geometry
assuming no upper-level drag ($\tau_{\rm drag}\to\infty$) and
$\tau_{\rm rad}$=10, 1, 0.1, and 0.01 Earth days from top to bottom, respectively.
Note the transition from a circulation dominated by zonally symmetric
jets at long $\tau_{\rm rad}$ to one dominated by day-to-night flow at
short $\tau_{\rm rad}$.}
\label{stswm-no-drag}
\end{minipage}
\end{figure}

The solutions confirm our theoretical predictions of a regime transition.
Figure~\ref{stswm-no-drag} illustrates the equilibrated solutions
for radiative time constants, $\tau_{\rm rad}$, of 10, 1, 0.1, 
and 0.01 days\footnote{In
this paper, 1 day is defined as $86400\sec$.} for the case where
drag is turned off (i.e., $\tau_{\rm drag}\to\infty$).\footnote{As
described by \citet{showman-polvani-2011}, the coupling between
layers---specifically, mass, momentum, and energy exchange in the presence
of heating/cooling---ensures that even cases without drag in the upper layer 
readily equilibrate to a steady state. All the models shown here are
equilibrated.}  As expected, when the
radiative time constant is long (10 days, panel (a)), jets dominate
the circulation, with relatively weak eddies in comparison to
the zonal-mean zonal winds.   At intermediate values of the radiative
time constant (1 and 0.1 days, panels (b) and (c)), the flow consists
of strong jets and superposed eddies.  At short values of the radiative
time constant (0.01 days, panel (d)), the jets are relatively weak---though
not absent---and day-to-night eddy flow dominates the circulation.

The dynamical behavior of this sequence can be understood as follows.
When the radiative time constant is long (Figure~\ref{stswm-no-drag}(a)),
the day-night thermal forcing is weak---air parcels experience only weak 
heating/cooling as they circulate from day to night---and the circulation 
is instead dominated by the equator-to-pole variation in the zonal-mean 
$h_{\rm eq}$ (i.e., by the zonal-mean radiative heating at low latitudes
and cooling at high latitudes).   At intermediate values of the
radiative time constant (panels (b) and (c)), the day-night thermal
forcing becomes sufficiently strong to generate a significant planetary
wave response, and the eddy-momentum convergence induced by
these waves generates equatorial superrotation via the mechanisms
identified by \citet{showman-polvani-2011}, particularly the differential
zonal propagation of the standing Kelvin and Rossby waves.  At short radiative
time constant (panel (d)), such zonal propagation is inhibited but,
as predicted by the theory in Section~\ref{theory}, the 
three-way force balance between pressure-gradient,
Coriolis, and advection forces still generates prograde phase tilts
in the velocities.  Although, visually, the flow appears dominated primarily
by day-to-night flow (Figure~\ref{stswm-no-drag}(d)), these phase tilts
still drive superrotation near the equator,
and even at higher latitudes the zonal-mean zonal wind remains a 
significant fraction of the eddy wind amplitude.

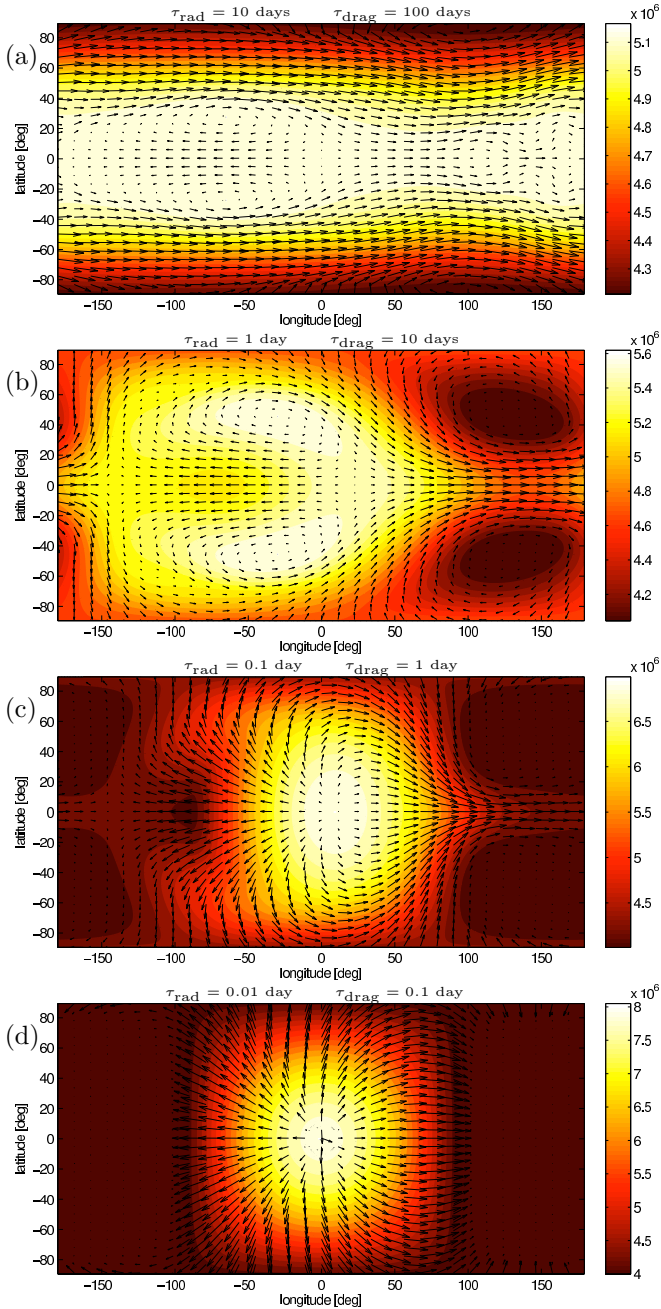
\begin{figure}
\begin{minipage}[c]{0.5\textwidth}
\includegraphics[scale=0.47, angle=0]{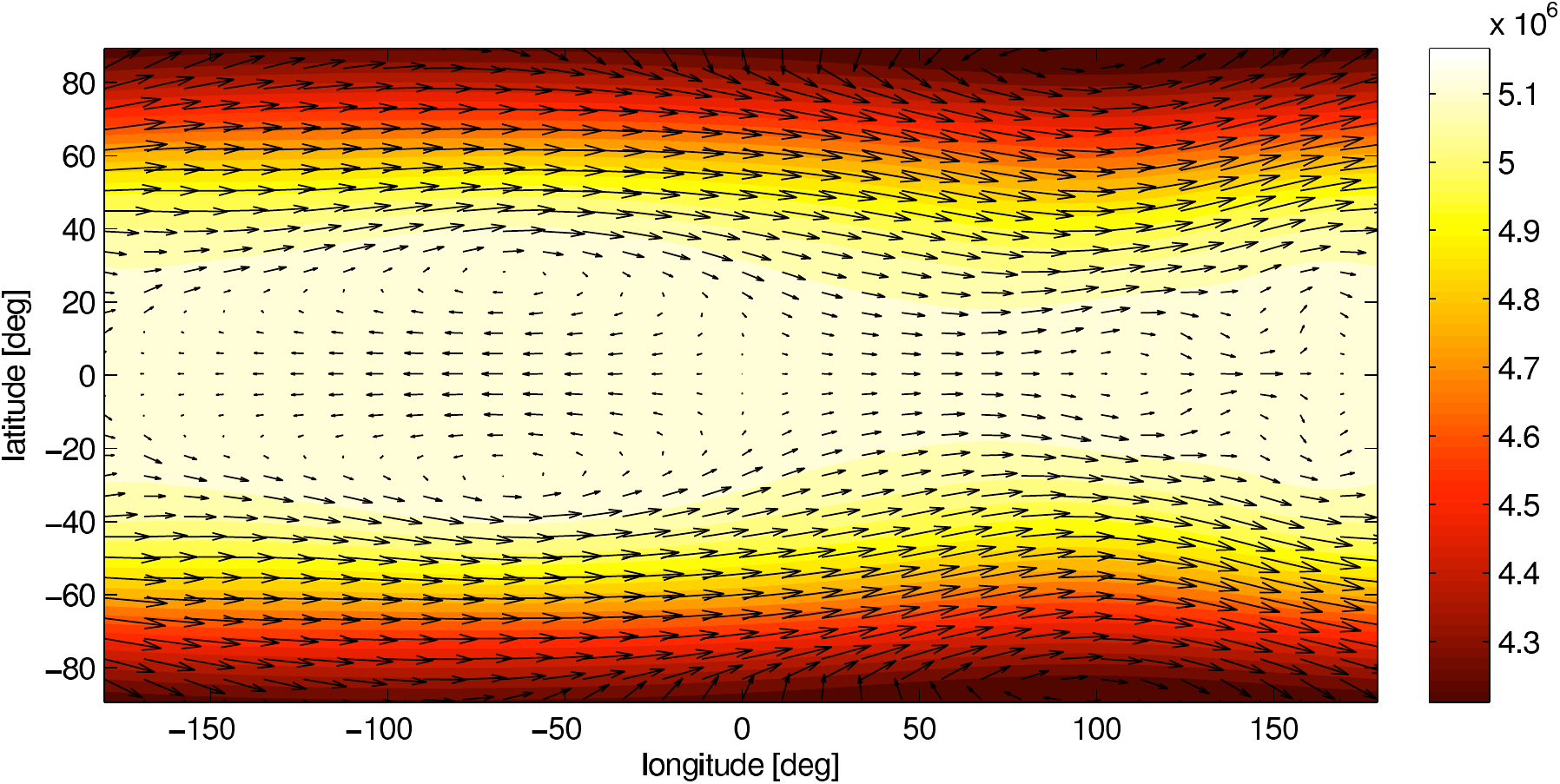}
\put(-251.,100.){\normalsize (a)}
\put(-185.,118.){\tiny $\tau_{\rm rad}=10$ days \qquad $\tau_{\rm drag}=100$ days}

\includegraphics[scale=0.47, angle=0]{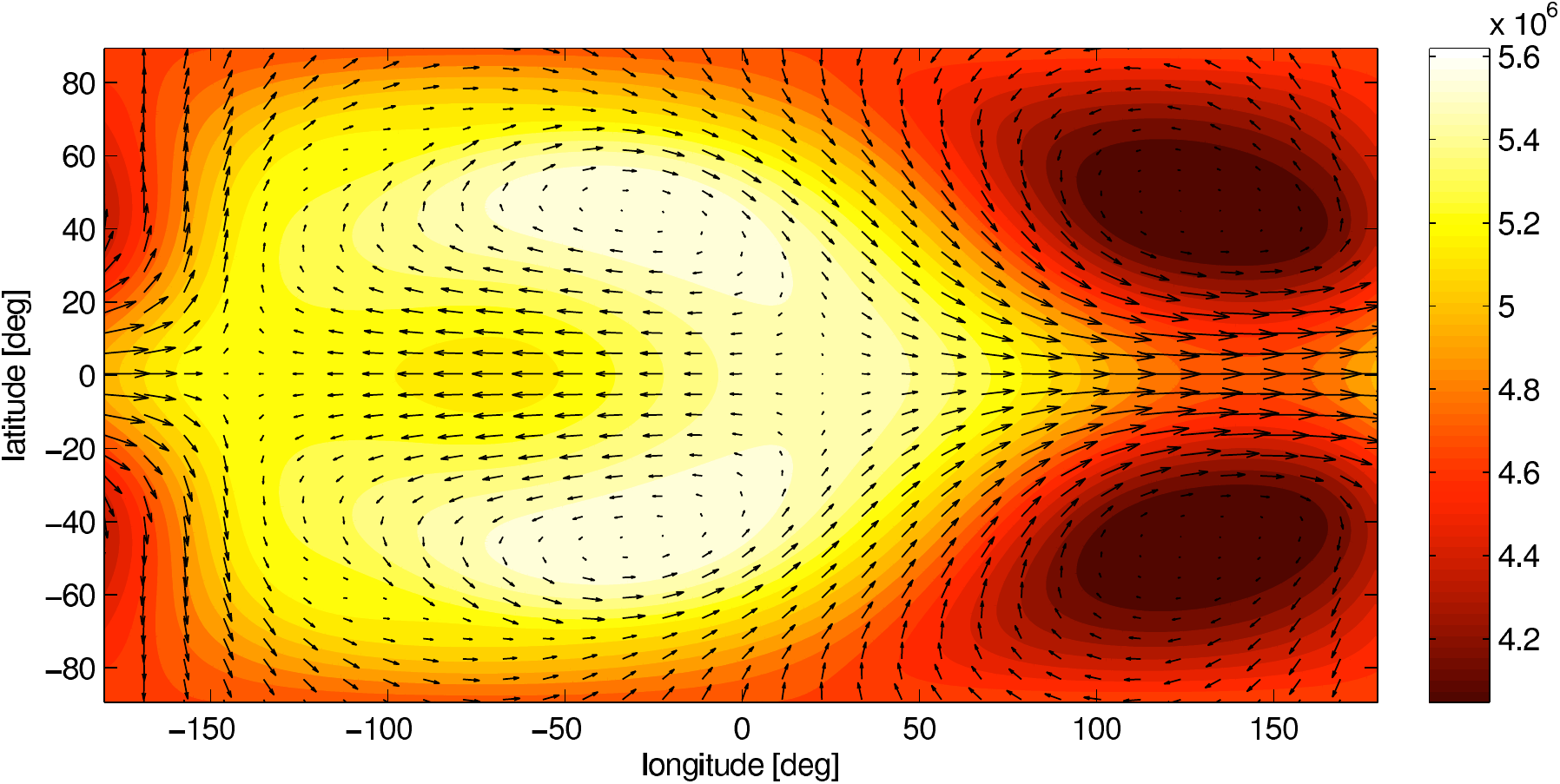}
\put(-251.,100.){\normalsize (b)}
\put(-180.,118.){\tiny $\tau_{\rm rad}=1$ day \qquad $\tau_{\rm drag}=10$ days}

\includegraphics[scale=0.47, angle=0]{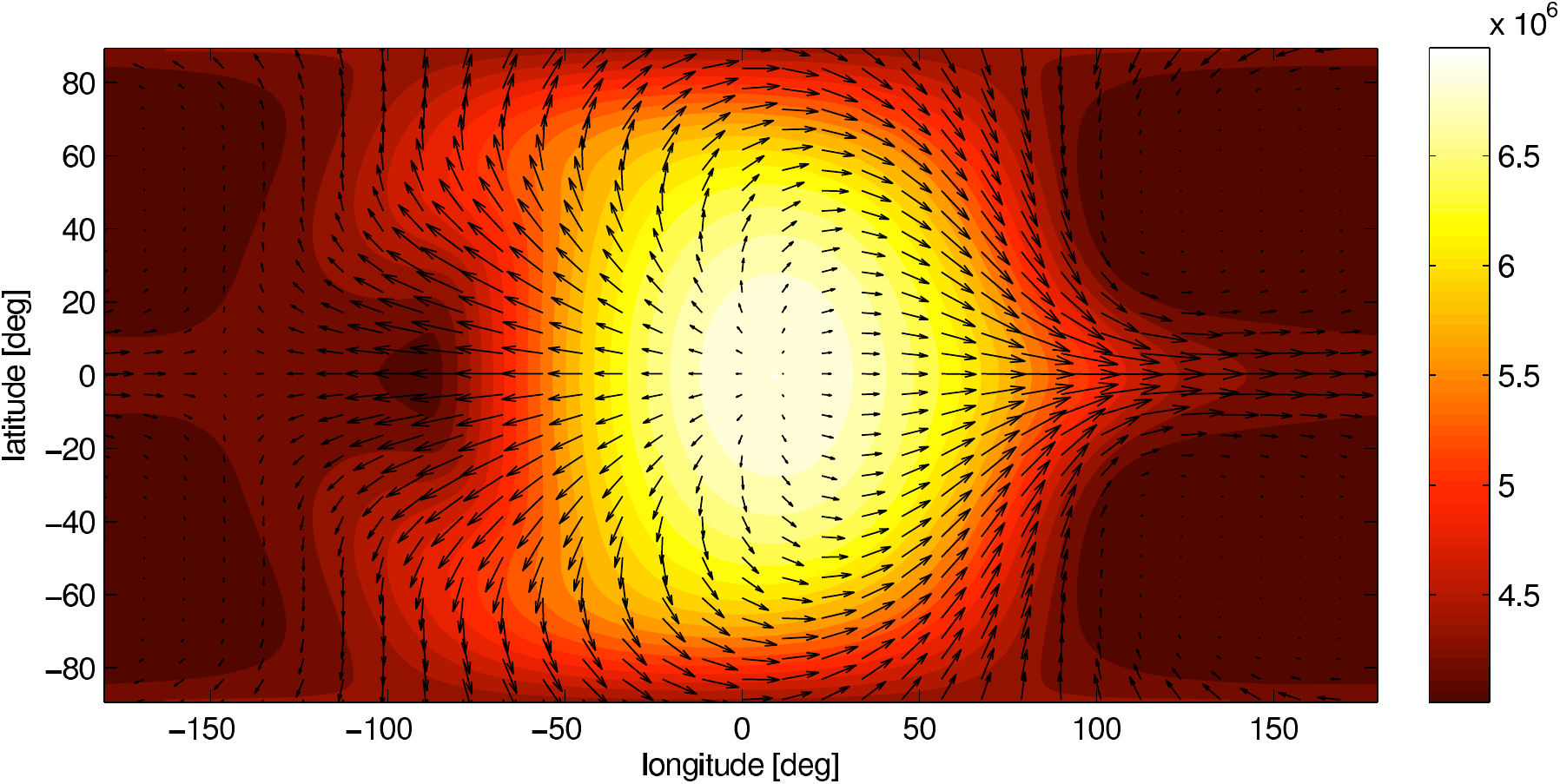}
\put(-251.,100.){\normalsize (c)}
\put(-180.,118.){\tiny $\tau_{\rm rad}=0.1$ day \qquad $\tau_{\rm drag}=1$ day}

\includegraphics[scale=0.47, angle=0]{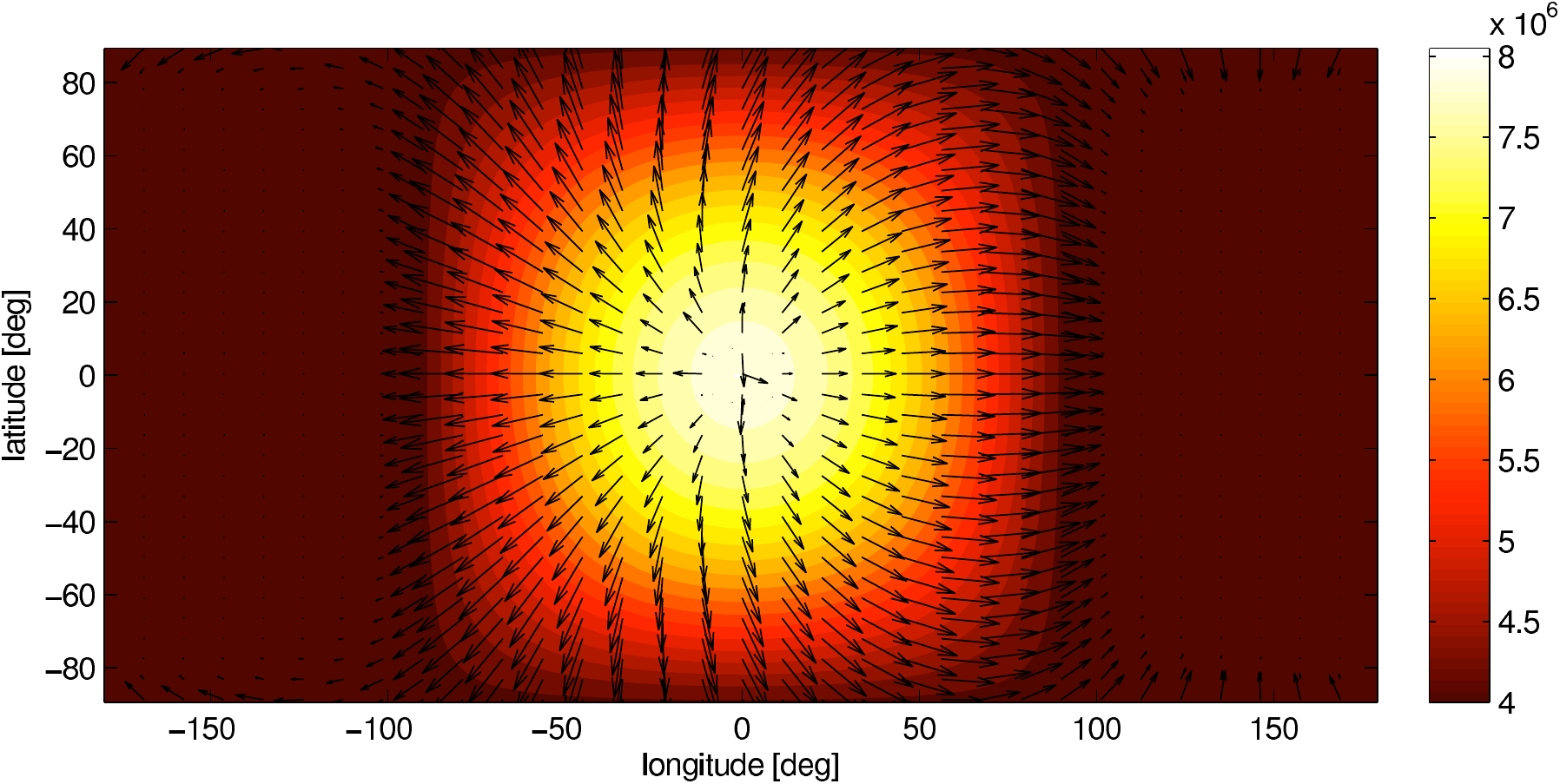}
\put(-251.,100.){\normalsize (d)}
\put(-187.,118.){\tiny $\tau_{\rm rad}=0.01$ day \qquad $\tau_{\rm drag}=0.1$ day}

\caption{Geopotential $gh$ (orange scale, units $\rm m^2\,s^{-2}$) and winds (arrows) for
the equilibrated (steady-state) solutions to the shallow-water equations
(Equations~\ref{momentum}--\ref{continuity}) in full spherical geometry
assuming $\tau_{\rm drag}=10\tau_{\rm rad}$ and
$\tau_{\rm rad}$=10, 1, 0.1, and 0.01 days from top to bottom, respectively.
Note the transition from a circulation dominated by zonally symmetric
jets at long $\tau_{\rm rad}$ to one dominated by day-to-night flow at
short $\tau_{\rm rad}$.}
\label{stswm-with-drag}
\end{minipage}
\end{figure}

The transition from a regime dominated by jets to a regime
dominated by day-to-night flow is even more striking when drag is
included.  Figure~\ref{stswm-with-drag} shows a sequence of models
with radiative time constants, $\tau_{\rm rad}$, of 10, 1, 0.1, and
0.01 days (as in Figure~\ref{stswm-no-drag}) but with $\tau_{\rm
  drag}=10\tau_{\rm rad}$ in all cases.  Overall, the trend resembles
that in Figure~\ref{stswm-no-drag}: at long radiative time constants
(panel (a)), the flow is dominated by high-latitude, highly zonal
jets; at intermediate radiative time constants (panel (b) and (c)),
the flow is transitional, exhibiting strong eddies associated with the
standing planetary-scale wave response to the day-night thermal
forcing, and zonal jets driven by those eddies
\citep[cf][]{showman-polvani-2011}; and at short radiative time
constants (panel (d)), the Kelvin and Rossby waves are damped and the
circulation consists almost entirely of day-to-night flow.  As
predicted by the theory in Section~\ref{theory}, drag in this case is
strong enough to overwhelm the advection and Coriolis forces, leading
to a two-way horizontal force balance between the pressure-gradient
force and drag.  As a result, there is no overall prograde phase tilt
of the velocity pattern. The eddy forcing of the zonal-mean flow,
and the jets themselves, are therefore weak.

To better characterize the dominance of jets versus day-night flow, 
we performed integrations over a complete grid including all
possible combinations of 0.01, 0.1, 1, 10, and 100 days in
$\tau_{\rm rad}$ and 0.1, 1, 10, 100, and $\infty$ days in $\tau_{\rm drag}$.
The typical amplitude of the jets can be characterized by the root-mean-square
of the zonal-mean zonal wind variation in latitude:
\begin{equation}
\overline{u}_{\rm rms} = \left[{1\over\pi}\int_{-\pi/2}^{\pi/2}
\overline{u}(\phi)^2 \,d\phi\right]^{1/2}
\end{equation}
where the overbar denotes a zonal average.  To characterize
the amplitude of the eddies, we adopt a metric representing the variation of
the zonal wind in longitude:
\begin{equation}
u_{\rm eddy}(\phi) =     \left[{1\over 2\pi}\int_{-\pi}^{\pi} 
(u-\overline{u})^2\,d\lambda\right]^{1/2}
\end{equation}
and then determine the root-mean-square variations of this quantity in latitude:
\begin{equation}
u_{\rm eddy, rms} = \left[{1\over\pi}\int_{-\pi/2}^{\pi/2}
u_{\rm eddy}(\phi)^2 \,d\phi\right]^{1/2}.
\end{equation}
The ratio of $\overline{u}_{\rm rms}$ to $u_{\rm eddy, rms}$ then
provies a measure of the relative dominance of jets versus 
day-night flow.

\begin{figure}
\includegraphics[scale=0.65, angle=0]{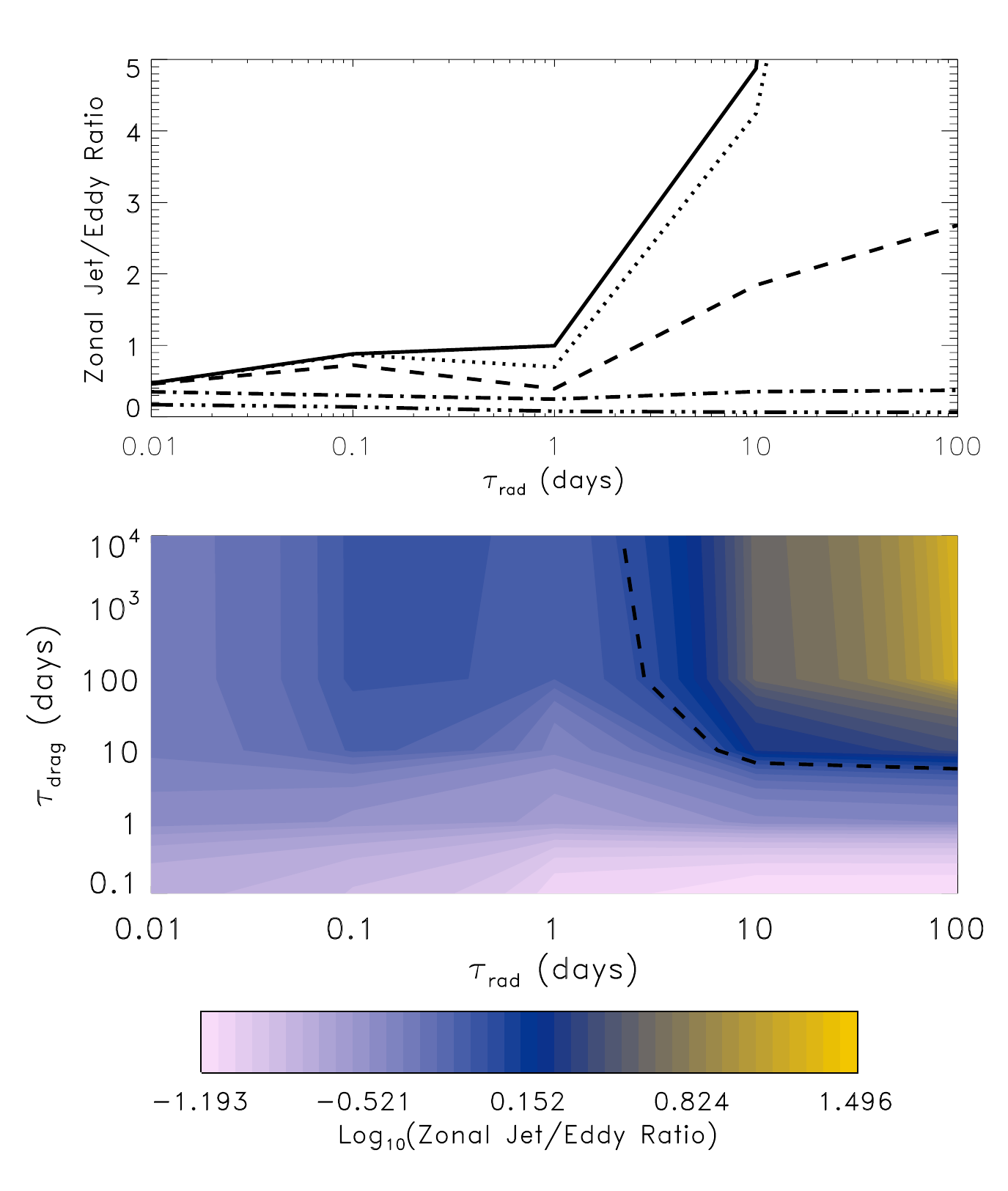}
\caption{Ratio $\overline{u}_{\rm rms}/u_{\rm eddy,rms}$, characterizing
the ratio of jets to eddies, versus $\tau_{\rm rad}$ and $\tau_{\rm drag}$.
{\it Top:} triple-dashed-dotted, dashed-dotted, dashed, dotted, and solid
curves depict results from models with 
$\tau_{\rm drag}$ values of 0.1, 1, 10, 100 days, and infinite
(i.e., no drag in the upper layer), respectively. {\it Bottom:} Two-dimensional
representation of the same data.  Colorscale depicts 
$\log_{\rm 10}(\overline{u}_{\rm rms}/u_{\rm eddy,rms})$ with even log-spacing
of the contour intervals. Values of $\overline{u}_{\rm rms}/u_{\rm eddy,rms}$
range from 0.06 (i.e., $10^{-1.193}$) to 31 (i.e., $10^{1.496}$).  The 
dashed black contour denotes $\overline{u}_{\rm rms}/u_{\rm eddy,rms}=1$
(i.e., $\log_{\rm 10}(\overline{u}_{\rm rms}/u_{\rm eddy,rms})=0$), giving
an approximate demarcation between the jet- and eddy-dominated regimes.
Jets dominate when both $\tau_{\rm rad}$ and $\tau_{\rm drag}$ are long,
and eddies (essentially day-night flow)
dominate when either time constant becomes very short.}
\label{jeteddy}
\end{figure}

These calculations demonstrate that jets dominate when both the
radiative and drag time constant are long, and that day-to-night
eddy flow dominates when either time constant is very short.
This is shown in Figure~\ref{jeteddy}, which 
presents the ratio $\overline{u}_{\rm rms}/u_{\rm eddy,rms}$
versus $\tau_{\rm rad}$ and $\tau_{\rm drag}$.   The dashed curve
in panel (b), corresponding to $\overline{u}_{\rm rms}/u_{\rm eddy,rms}=1$, demarcates
the approximate transition between regimes (jets dominate above and to
the right of the curve, while day-night flow dominates below and to the
left of the curve).  Although extremely short values of {\it either}
$\tau_{\rm rad}$ or $\tau_{\rm drag}$ are sufficient to ensure 
eddy-dominated flow, the trend of the transition differs for the
two time constants.  When drag is weak or absent, $\tau_{\rm rad}$ must be
extremely short---less than 0.1 day---to ensure eddy- rather than jet-dominated
flow (Figure~\ref{jeteddy}).  
On the other hand, over a wide range of $\tau_{\rm rad}$ values,
$\tau_{\rm drag}$ need only be less than $\sim$3 days to ensure eddy-dominated
flow.  The transition between jet and eddy-dominated regimes as a
function of drag occurs more sharply when $\tau_{\rm rad}$ is large
than when it is small.

\section{Three-dimensional models}
\label{3D}

We now demonstrate this regime shift in 
three-dimensional atmospheric circulation models including realistic
radiative transfer.  GJ 436b, HD 189733b, and HD 209458b are selected
as examples that bracket a large range in stellar irradiation
and yet are relatively easily observable, hence representing
good targets for Doppler characterization.

\subsection{Model setup}

We solve the radiation hydrodynamics equations using the Substellar
and Planetary Atmospheric Circulation and Radiation (SPARC) model of
\citet{showman-etal-2009}.  This model couples the dynamical core of
the MITgcm \citep{adcroft-etal-2004}, which solves the primitive
equations of meteorology in global, spherical geometry, using pressure
as a vertical coordinate, to the state-of-the-art, nongray radiative
transfer scheme of \citet{marley-mckay-1999}, which solves the
multi-stream radiative transfer equations using the correlated-$k$
method to treat the wavelength dependence of the opacities.  Here, we
use a two-stream implementation of this model.  To date, this is the
only circulation model of hot Jupiters to include a realistic
radiative transfer solver, which is necessary for accurate
determination of heating rates, temperatures, and flowfield.  The
composition and therefore opacities in hot-Jupiter atmospheres are
uncertain.  Here, gaseous opacities are calculated assuming local
chemical equilibrium for a specified atmospheric metallicity, allowing
for rainout of any condensates.  We neglect any opacity due to clouds
or hazes.

Model parameters are summarized in Table~\ref{params}.  Although the
atmosphere of GJ 436b is likely enriched in heavy elements
\citep{spiegel-etal-2010, lewis-etal-2010, madhusudhan-seager-2011},
solar metallicity represents a reasonable baseline for HD 189733b and
HD~209458b, and to establish the effect of differing stellar flux at
constant metallicity we therefore adopt solar metallicity for the gas
opacities in our nominal models of all three planets.  To bracket the
range of plausible metallicities, we also explore a model of GJ 436b
with 50 times solar metallicity. For HD 189733b and GJ 436b, we
neglect opacity due to strong visible-wavelength absorbers such as TiO
and VO, as TiO and VO are not expected for these cooler planets.  For
HD 209458b, secondary-eclipse measurements suggest the presence of a
stratosphere \citep{knutson-etal-2008}, and we therefore include TiO
and VO opacity for this planet, which allows a thermal inversion due
to the large visible-wavelength opacity of these species
\citep{hubeny-etal-2003, fortney-etal-2008, showman-etal-2009}.  While debate exists
about the ability of TiO to remain in the atmosphere \citep{showman-etal-2009, 
spiegel-etal-2009}, our present purpose is simply
to use TiO as a proxy for any chemical species that strongly absorbs
in the visible wavelengths and hence allows a stratosphere to exist;
other strong visible-wavelength absorbers would exert qualitatively
similar effects.  Synchronous rotation is
assumed for HD 189733b and HD 209458b (with a substellar longitude
perpetually at $0^{\circ}$); the rotation of GJ 436b, however, is
assumed to be pseudosychronized with its slightly eccentric orbit
\citep{lewis-etal-2010}.\footnote{This has only a modest effect
on the results; synchronously rotating models of GJ 436b on circular
orbits exhibit similar circulation patterns.} 
The obliquity of all models is zero so that
the substellar point lies along the equator.

\begin{deluxetable}{cccccc}
\tabletypesize{\tiny}
\tablecolumns{4}
\tablewidth{0pc}
\tablecaption{Model Parameters}
\tablehead{
\colhead{} & \colhead{GJ 436b} & \colhead{HD 189733b} & \colhead{HD 209458b}}
\startdata

   radius (m) & $2.69\times10^7$ &$8.2396\times10^7$  & $9.437\times10^7$ \\
   rotation period (days) &2.3285 &2.2 &3.5\\
   gravity ($\rm m\,s^{-2}$) &12.79 &21.4 &9.36\\
   $p_{\rm base} (\rm bars)$ &200 & 200 & 200\\
   $p_{\rm top} (\rm bars)$ &$2\times10^{-5}$ &$2\times10^{-6}$ &$2\times10^{-6}$\\
   $N_r$ &47 &53 &53\\
   metallicity (solar) &1 and 50  &1  &1\\
\enddata
\label{params}
\end{deluxetable}

Our nominal models do not include explicit frictional drag in the
upper levels.\footnote{All of the models include a Shapiro filter to
maintain numerical stability.  In some of the models, particularly those
for HD 209458b, we also include a drag term in the deep atmosphere
below 10 bars; this allows the total kinetic energy of the model to
equilibrate while minimally affecting the circulation in the upper atmosphere.
Note that, for computationally feasible integration times (thousands of
Earth days), models that include drag in the deep layers (but not at pressures
less than $\sim$10 bars) exhibit flow 
patterns and wind speeds in the observable atmosphere that are extremely
similar to those in models that entirely lack large-scale drag.  This is due
to the fact that, even in such drag-free models, the wind speeds at pressures
$\gtrsim 10\,$bars remain weak.  For brevity, in this paper, we use the term
``drag free'' to refer to models lacking an explicit large-scale drag term, 
$-{\bf v}/\tau_{\rm drag}$,  in the observable atmosphere; nevertheless, it should be
borne in mind that some of those models do contain drag in the bottommost
model layers, and all of them include the Shapiro filter. }  However, several frictional
processes may be important for hot Jupiters, including vertical
turbulent mixing \citep{li-goodman-2010}, breaking small-scale gravity
waves \citep{watkins-cho-2010}, and magnetohydrodynamic drag
\citep{perna-etal-2010}.  The latter may be particularly important for
hot planets such as HD 209458b.  Accordingly, we additionally present
a sequence of HD 209458b integrations that include frictional drag,
which we crudely parameterize as a linear relaxation of the zonal and
meridional velocities toward zero\footnote{In other words, we add a
  term $-{\bf v}/\tau_{\rm drag}$ to the horizontal momentum
  equations, where ${\bf v}$ is the horizontal velocity.  This simple
  scheme is called ``Rayleigh drag'' in the atmospheric dynamics
  literature.} over a prescribed drag time constant, $\tau_{\rm
  drag}$.  Within any given model, we treat $\tau_{\rm drag}$ as a
constant everywhere within the domain.
This is not a rigorous representation of drag (for example,
Lorentz forces will vary greatly from dayside to nightside and may act
anisotropically on the zonal and meridional winds); still, the
approach allows a straightforward evaluation of how drag of a given
strength alters the circulation.

For all three planets, we solve the equations on the cubed-sphere grid
using a horizontal resolution of C32, corresponding to an approximate
global resolution of $128\times 64$ in longitude and latitude.  The
lowermost $N_r-1$ vertical levels are evenly spaced in log-pressure from an
average basal pressure $p_{\rm base}$ of 200 bars to a top pressure, 
$p_{\rm top}$, of $20\,\mu$bars for GJ 436b and $2\,\mu$bars for 
HD 189733b and HD 209458b.  The uppermost model level extends from a pressure
of $p_{\rm top}$ to zero.  Our
models of GJ 436b and HD 189733b were originally presented in
\citet{lewis-etal-2010} and \citet{fortney-etal-2010}, respectively,
while for HD 204958b we present new models here.  These new
integrations adopt 11 opacity bins in our correlated-$k$ scheme;
detailed tests show that this 11-bin scheme produces net radiative fluxes,
heating rates, and atmospheric circulations very similar to those
of the 30-bin models
\citep[see][]{kataria-etal-2012}. We integrate these models until
they reach an essentially steady flow configuration at pressures $<1\,$bar,
corresponding to integration times typically ranging from one to four thousand
Earth days, depending on the model.

\subsection{Results: nominal models}

\begin{figure}
\begin{minipage}[c]{0.5\textwidth}
\includegraphics[scale=0.45, angle=0]{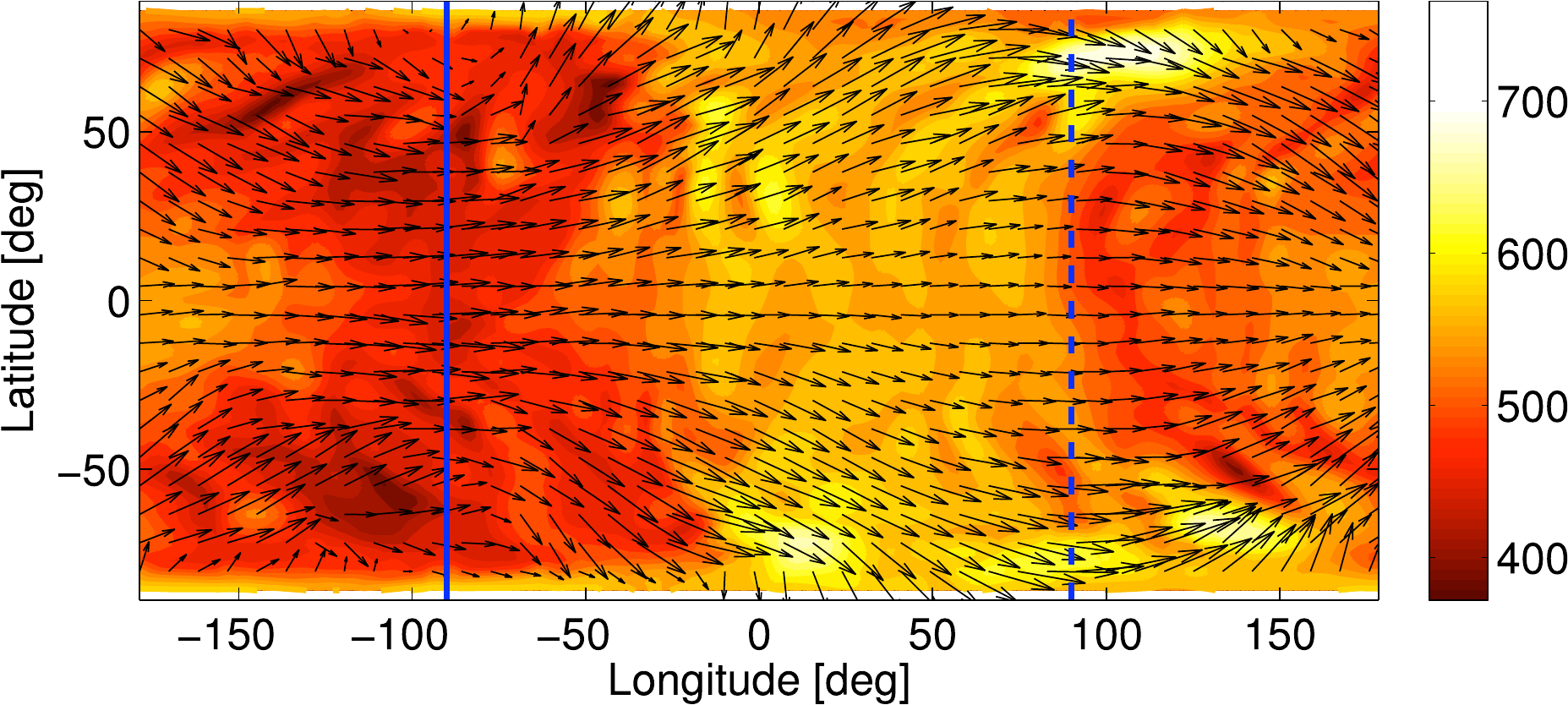}
\put(-243.,100.){\large (a)}\\
\includegraphics[scale=0.45, angle=0]{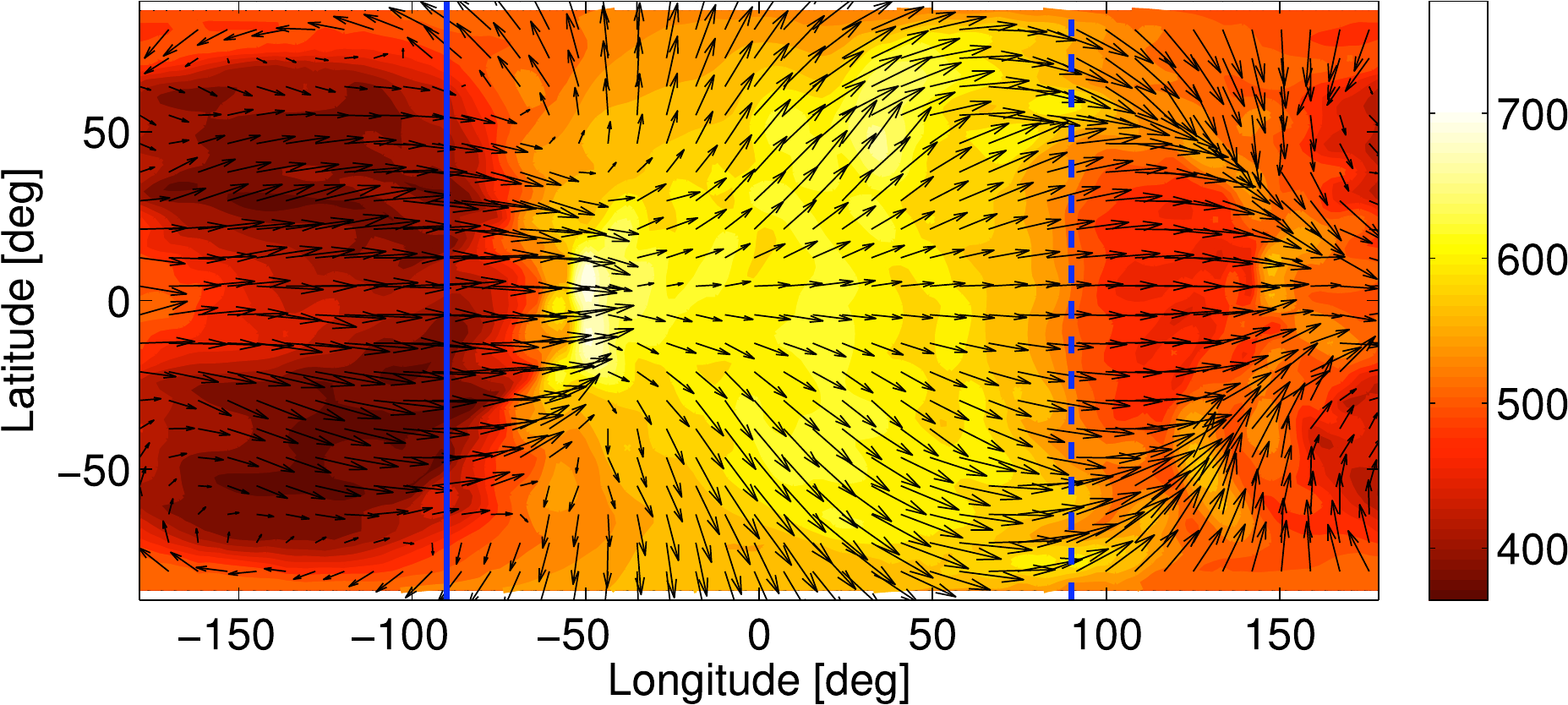}
\put(-243.,100.){\large (b)}\\
\includegraphics[scale=0.45, angle=0]{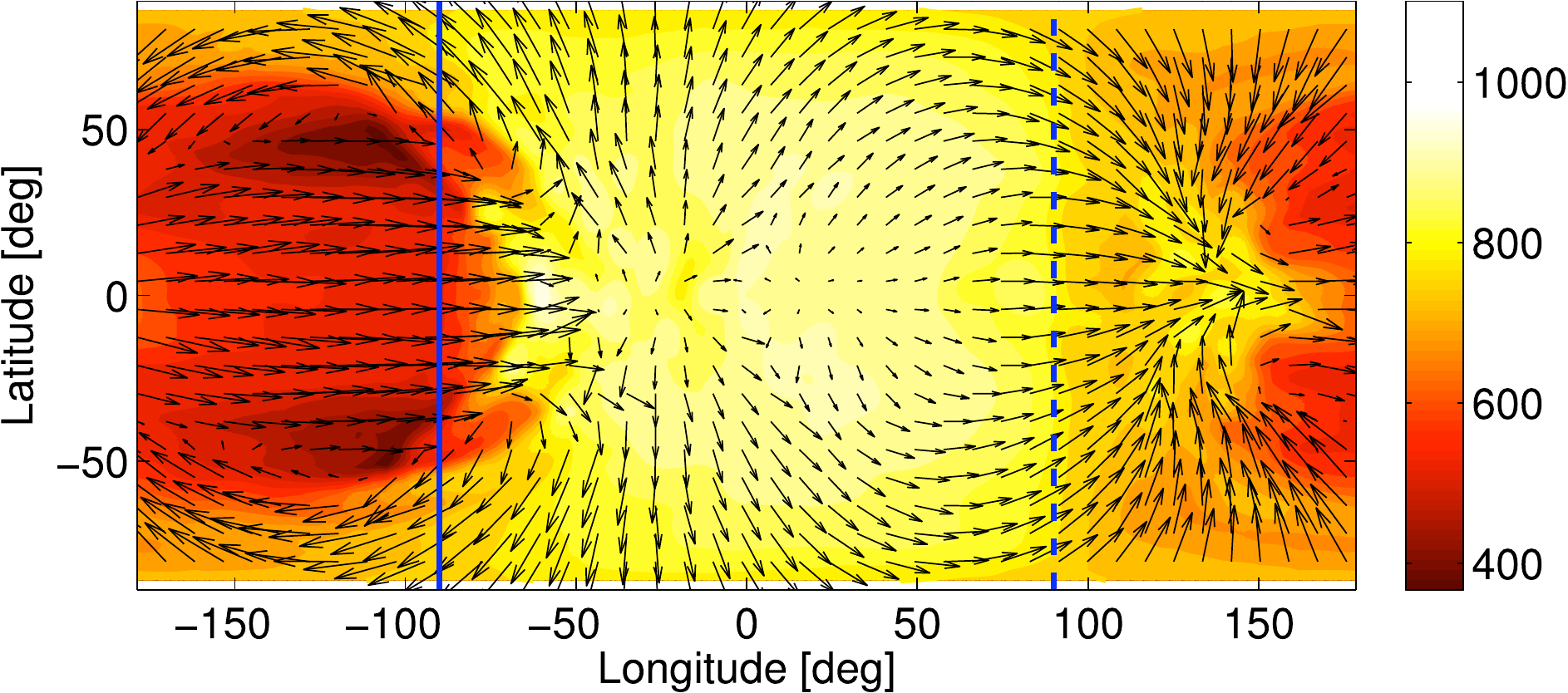}
\put(-243.,100.){\large (c)}\\
\includegraphics[scale=0.45, angle=0]{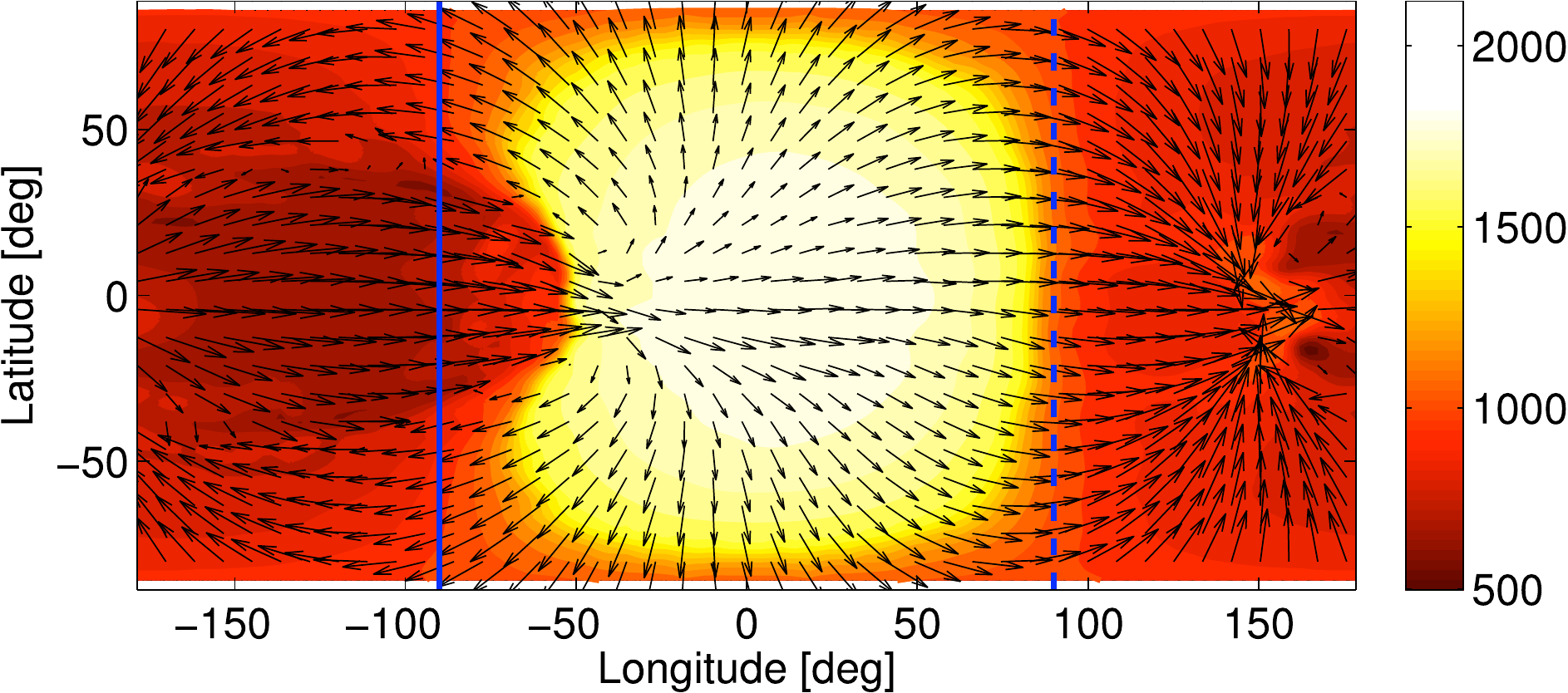}
\put(-243.,100.){\large (d)}\\
\caption{Global temperature (orange scale) and winds (arrows) for
a sequence of 3D SPARC/MITgcm
models at a pressure of 0.1 mbar, where Doppler measurements are 
likely to sense.  Panels (a), (b), (c), and (d) show solar-metallicity
GJ 436b, $50\times$ solar GJ 436b, solar-metallicity HD 189733b, and 
solar-metallicity HD 209458b models, respectively.
The vertical blue solid and dashed lines show the
location of the terminators $90^{\circ}$ west and east of the substellar longitude,
respectively.  The substellar point
is at longitude $0^{\circ}$ in all panels.  The models show a gradual transition
from a circulation dominated by zonal jets (top) to one dominated by
day-night flow at low pressure (bottom).}
\label{temp-winds}  
\end{minipage}
\end{figure}

\begin{figure}
\begin{minipage}[c]{0.5\textwidth}
\includegraphics[scale=0.45, angle=0]{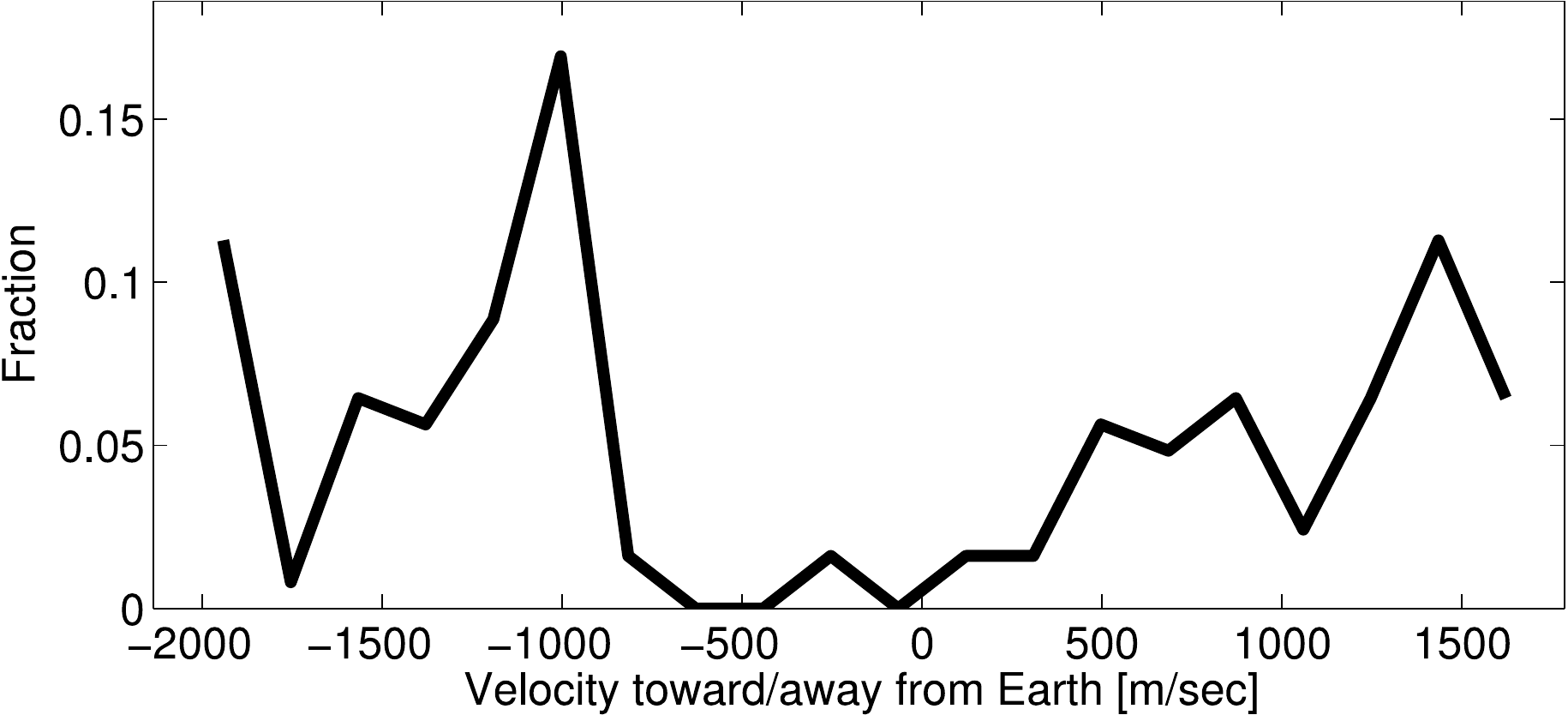}
\put(-243.,100.){\large (a)}\\
\includegraphics[scale=0.45, angle=0]{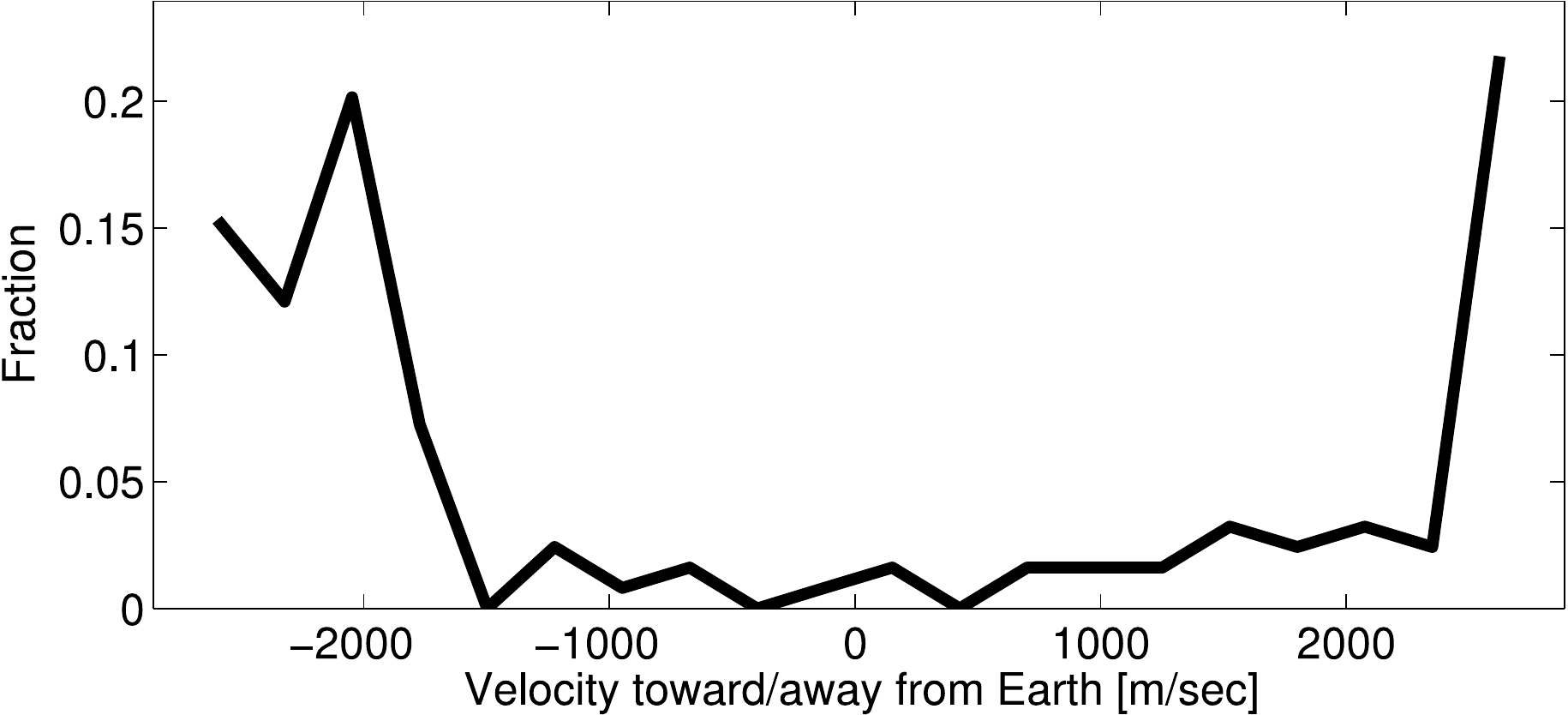}
\put(-243.,100.){\large (b)}\\
\includegraphics[scale=0.45, angle=0]{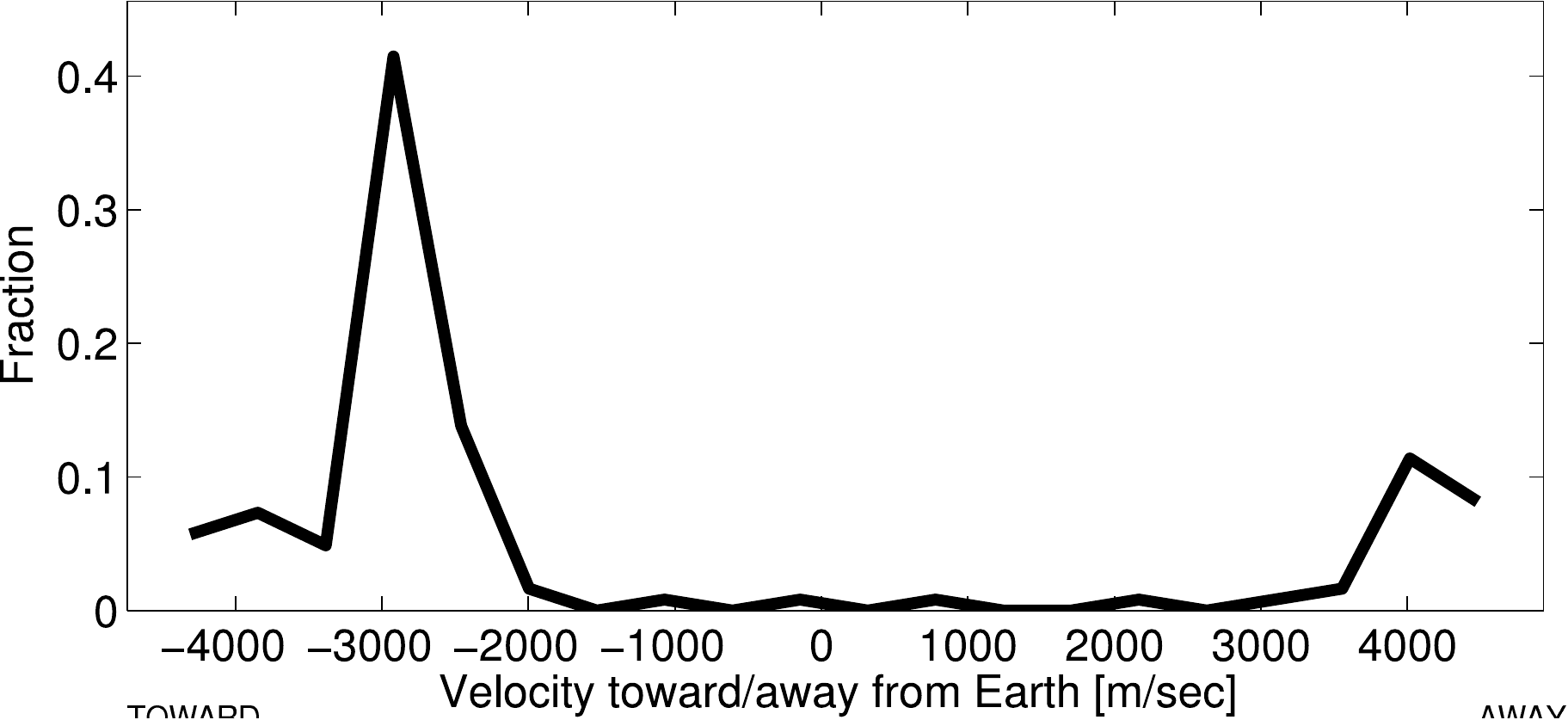}
\put(-243.,100.){\large (c)}\\
\includegraphics[scale=0.45, angle=0]{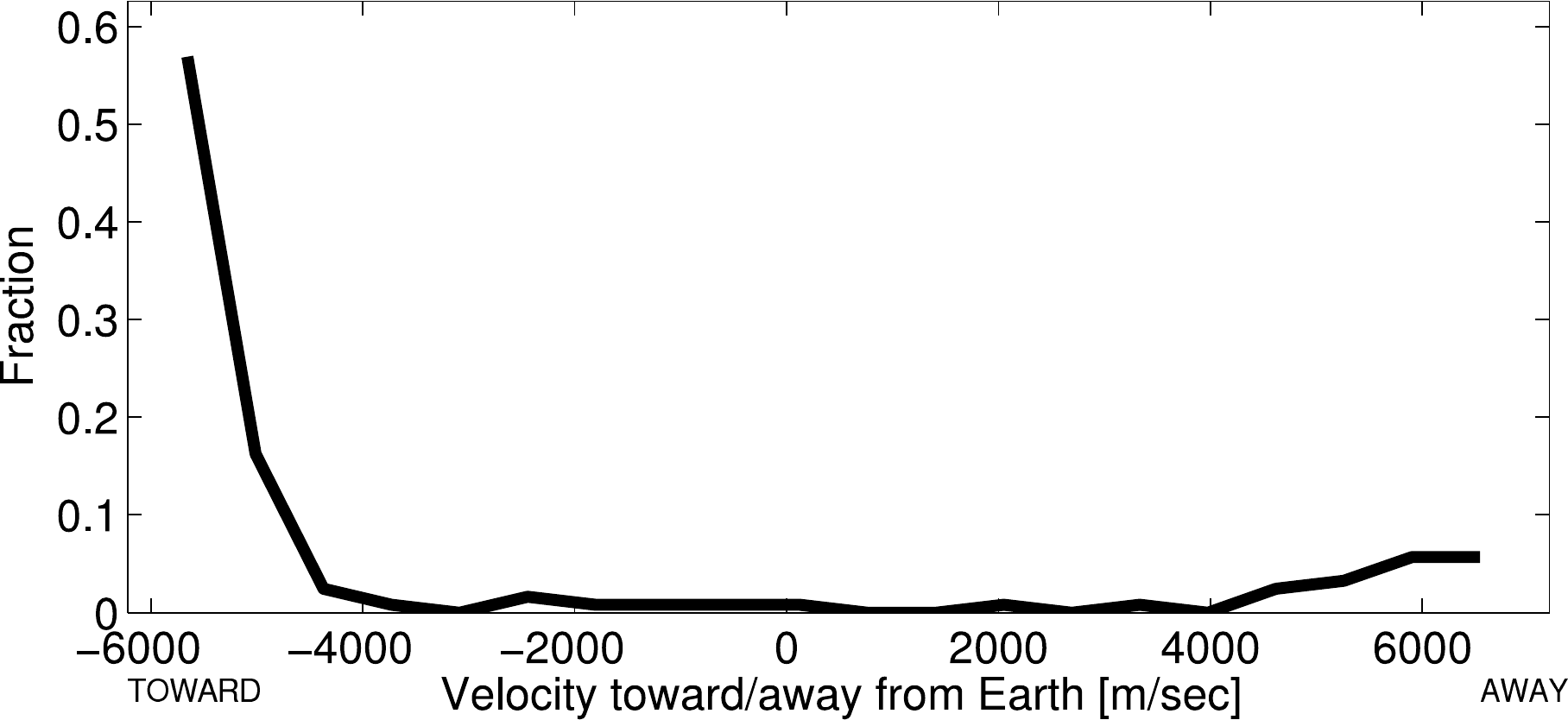}
\put(-243.,100.){\large (d)}

\caption{Histograms showing the fraction of the full, $360^{\circ}$ terminator,
at a pressure of 0.1 mbar, flowing at various wind speeds toward or away from Earth 
for the four models presented
in Figure~\ref{temp-winds}.   (To isolate the dynamical contribution,
this does not include the contribution of planetary rotation to the
inertial-frame velocity.)  Negatives values are toward Earth and positive
values are away from Earth.  Panels (a), (b), (c), and (d) show solar-metallicity
GJ 436b, $50\times$ solar GJ 436b, solar-metallicity HD 189733b, and 
solar-metallicity HD 209458b models, respectively. This gives a crude sense of
the dynamical contribution to the 
Doppler splitting of a discrete spectral line.  The models exhibit a transition
from flows that exhibit both blue- and red-shifted components 
(top) to a flow whose Doppler
signature would be predominantly blueshifted (bottom).}
\label{histogram}
\end{minipage}
\end{figure}

Our nominal, three-dimensional models exhibit a fundamental transition
in the upper-atmospheric behavior---at pressures where Doppler
measurements are likely to sense---as stellar insolation increases
from modest (for GJ 436b) to intermediate (for HD 189733b) to large
(for HD 209458b).  This is illustrated in Figures~\ref{temp-winds} and
\ref{histogram}.  Figure~\ref{temp-winds} shows temperature and winds
over the globe for models of GJ 436b, HD 189733b, and HD
209458b. Figure~\ref{histogram} presents histograms of the fraction of
terminator arc length versus terminator zonal-wind speed for these
same models,\footnote{For each model, we define a one-dimensional 
array $u_i$ corresponding to the terminator velocity at 0.1 mbar versus terminator
angle $\theta_i$ from 0 to $360^{\circ}$.   We define 20 velocity
bins, equally spaced between the minimum and maximum terminator velocities
from the array $u_i$. 
We then determine the fraction of the points in the $u_i$ array that
fall into each velocity bin.  This is what is plotted in Figure~\ref{histogram}.
The qualitative results are similar
when different choices are made for the number of velocity bins.} which
gives an approximate sense of how a discrete spectral line would be
split, shifted, or smeared in frequency due to the Doppler shift of
zonal winds along the terminator.  To isolate the effect of dynamics,
the contribution of planetary rotation to the velocity is not included in 
Figure~\ref{histogram}, although we will return to its effects
subsequently.

In the case of solar-metallicity GJ 436b (Figure~\ref{temp-winds}a),
the predominant dynamical feature is a broad superrotating (eastward)
jet that extends over all longitudes and in latitude almost from pole
to pole.  The jet exhibits significant wave activity, manifesting as
small-scale fluctuations in temperature and zonal wind, particularly
at the high latitudes of both hemispheres where the zonal-mean zonal
winds peak.  Nevertheless, the model exhibits little tendency toward a
zonal-wavenumber-one pattern that would be associated with a
predominant day-to-night flow.  Save for small regions near the poles,
the zonal winds at low pressure are everywhere eastward, implying that, during
transit, the zonal winds flow away from Earth along the leading limb
and toward Earth along the trailing limb.  This would lead to almost
equal blueshifted and redshifted Doppler components, with a relative
minimum near zero Doppler shift (Figure~\ref{histogram}a).

Next consider HD 189733b (Figure~\ref{temp-winds}c).  The model again
exhibits a superrotating equatorial jet, which is fast across most of
the nightside---achieving eastward speeds of $4\rm\,km\,s^{-1}$---but
slows down considerably over the dayside, reaching zero speed near the
substellar point.  Despite this variation, the zonal winds within the
jet (latitudes equatorward of $60^{\circ}$) are eastward along both
terminators.  In contrast, the high-latitude zonal wind (poleward of
$60^{\circ}$ latitude) is westward along the western terminator and
eastward along the eastern terminator\footnote{Eastern and western
  terminators refer here to the terminators $90^{\circ}$ of longitude
  east and west, respectively, of the substellar point.}, as expected
for day-to-night flow.  As seen during transit, along the trailing
limb, the zonal winds flow toward Earth.  Along the leading limb, they
flow toward Earth poleward of $60^{\circ}$ latitude and away from
Earth equatorward of $60^{\circ}$ latitude.  Spectral lines would thus
exhibit a broadened or bimodal character, with the blueshifted component
considerably stronger than the redshifted component
(Figure~\ref{histogram}c).  HD 189733b is thus a transitional case
between the two regimes discussed in Section~\ref{theory}.

In the case of HD 209458b (Figure~\ref{temp-winds}d), the strong
superrotating jet continues to exist, but at and west of the western terminator 
it is confined substantially closer to the equator than in our GJ 436b or 
HD 189733b models.   Poleward of $\sim$$30^{\circ}$ latitude on the
western terminator, and everywhere along the eastern terminator, the
airflow direction is from day to night.  This implies that, as seen
during transit, the trailing limb exhibits zonal winds toward
Earth.  The leading limb exhibits zonal winds that are toward Earth
poleward of $\sim$$30^{\circ}$ latitude and away from Earth equatorward
of $\sim$$30^{\circ}$ latitude.  This would lead to Doppler shifts
that are almost entirely blueshifted (Figure~\ref{histogram}d).

\begin{figure}
\includegraphics[scale=0.45, angle=0]{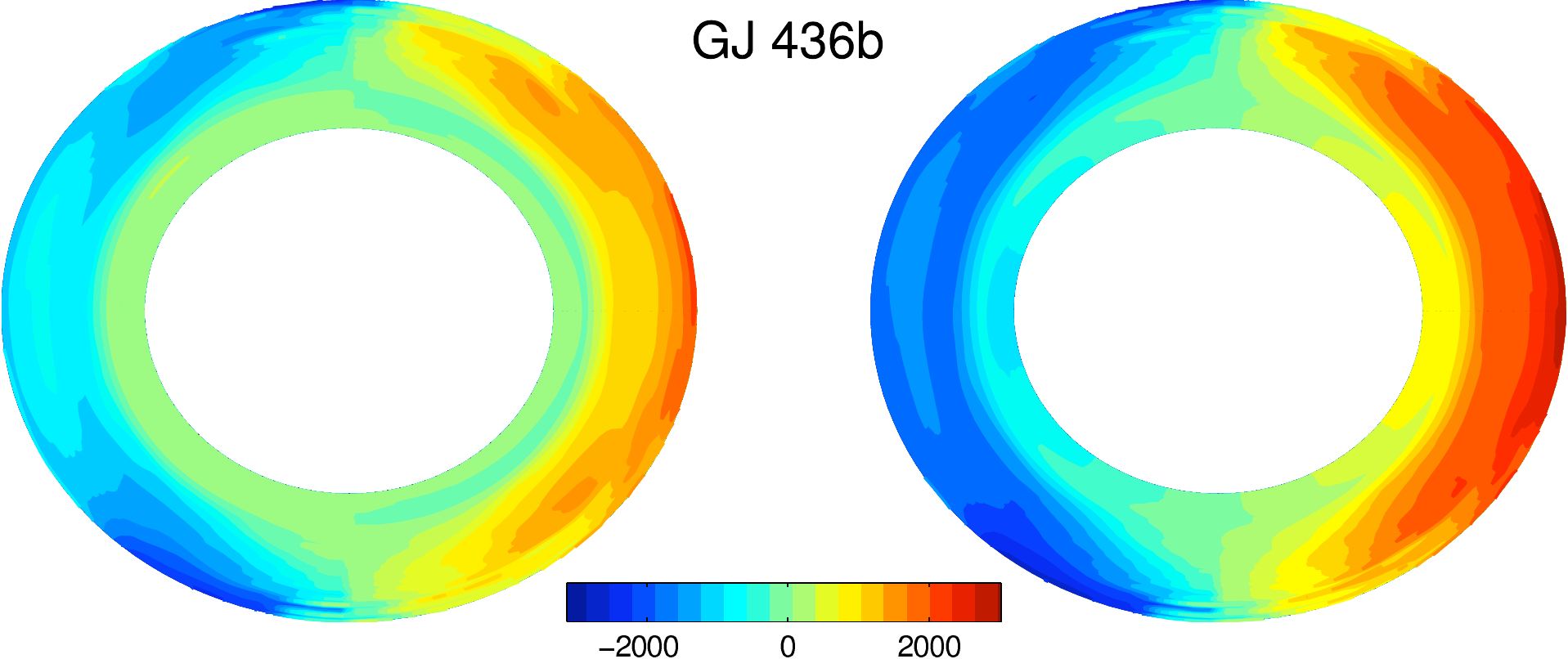}\\
\vskip 5pt
\includegraphics[scale=0.45, angle=0]{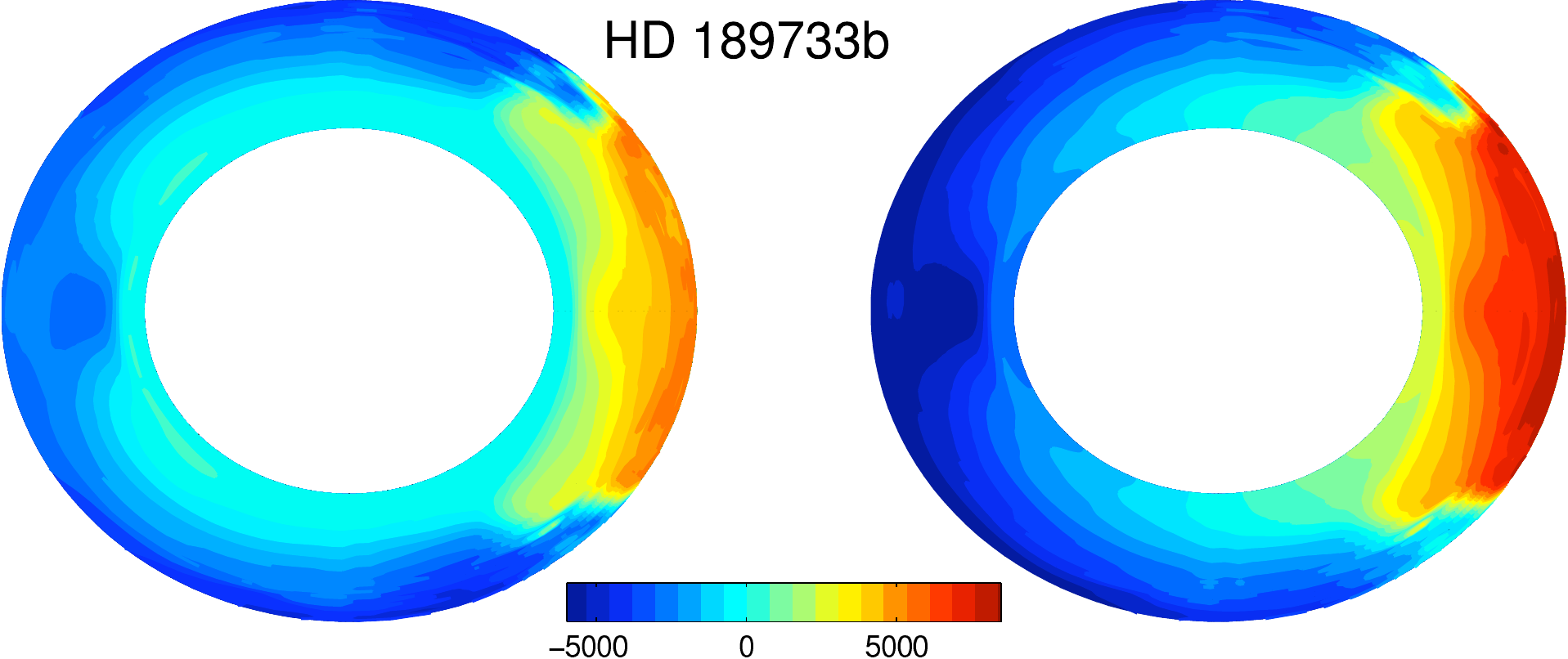}\\
\vskip 5pt
\includegraphics[scale=0.45, angle=0]{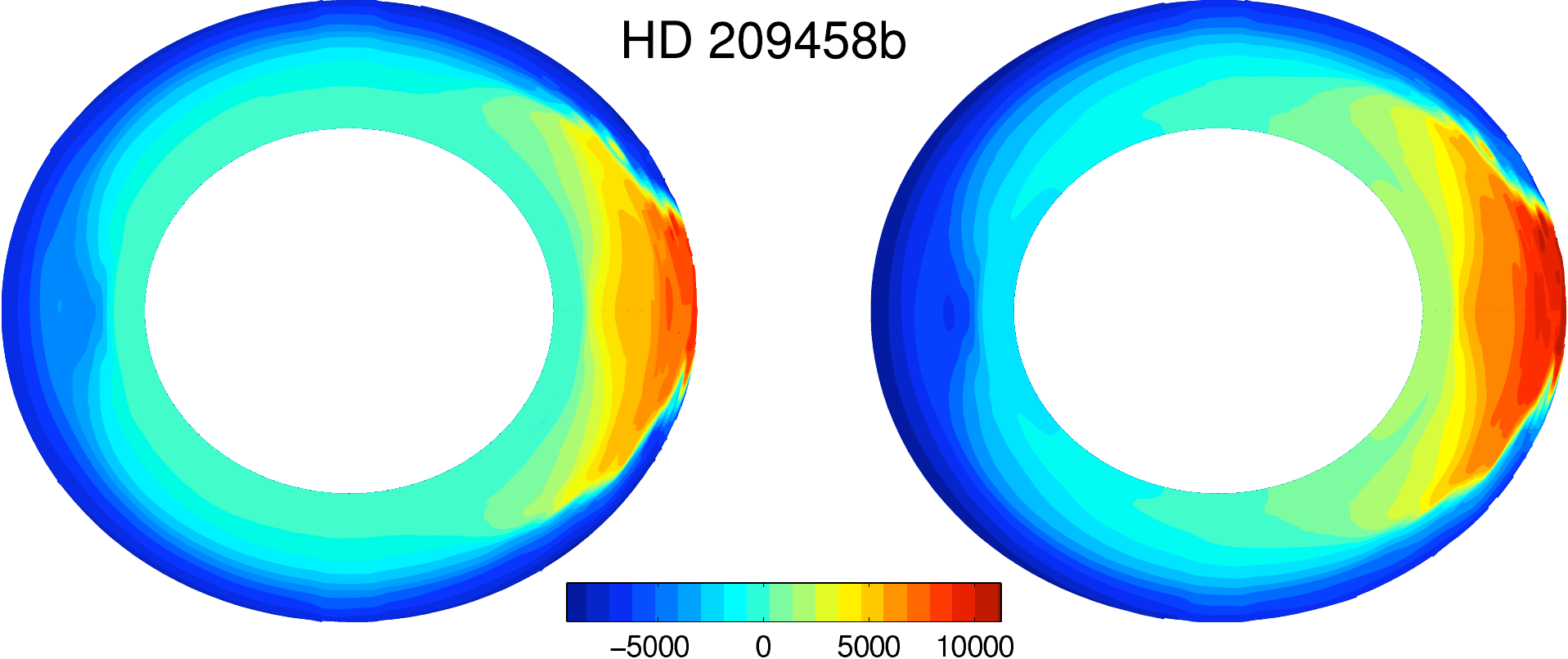}\\
\vskip 5pt
\includegraphics[scale=0.45, angle=0]{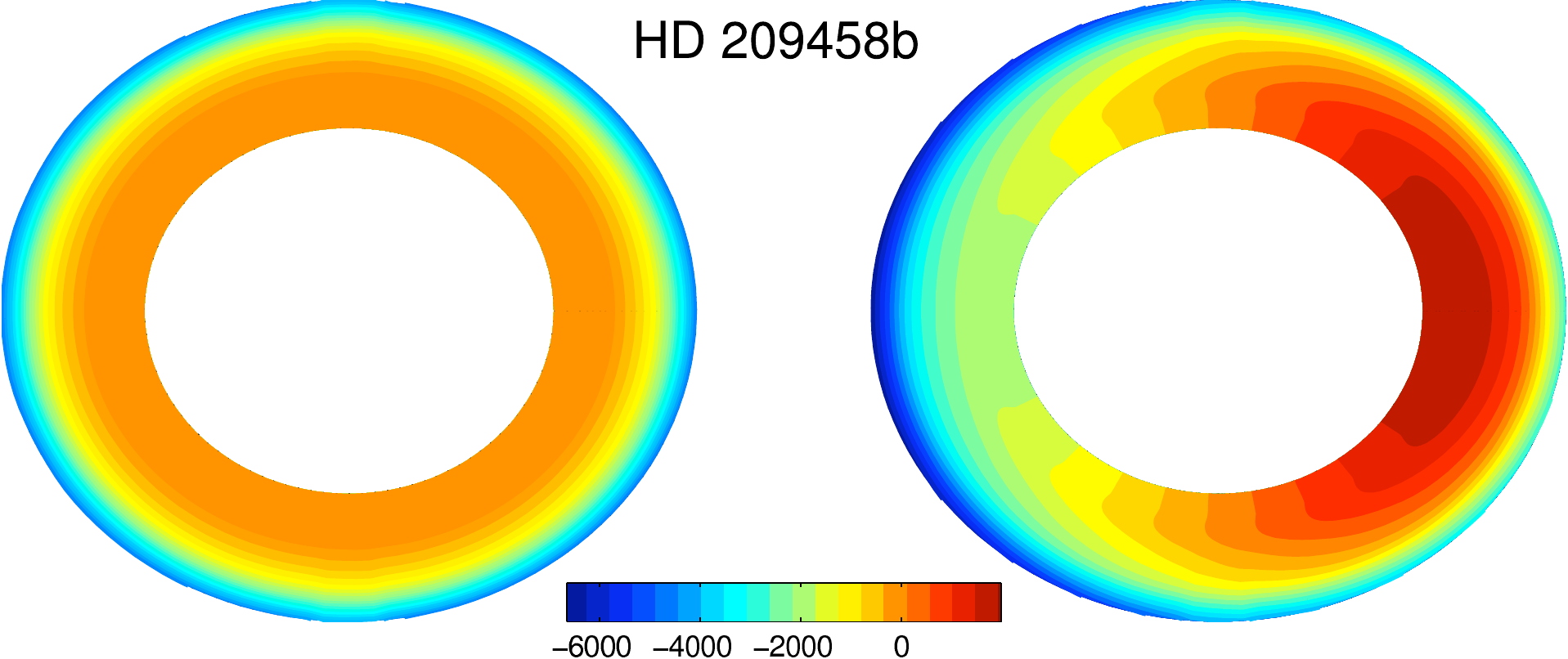}
\caption{Winds toward or away from Earth 
(colorscale, $\rm m\,s^{-1}$) along the full, $360^{\circ}$ terminator
in a sequence of models as viewed during the center of transit.  Colorscale
is such that red (positive) represents redshifted velocities while blue
(negative) represents blueshifted velocities.  The radial coordinate represents 
log pressure, and the plotted range is from 200 bars at the inside to
$2\,\mu$bar at the outside. The first, second, and
third rows show our solar-metallicity nominal models of GJ 436b, HD 189733b,
and HD 209458b, respectively.  The fourth row shows our model of 
HD 209458b where frictional drag is imposed with a drag time constant
of $10^4\rm\,s$.  For each model, the left panel shows the winds alone,
and the right panel shows the sum of the winds and the planet's rotation.
From top to bottom, the transition from high-altitude velocities that have 
both blue and redshifted components to velocities that are entirely blueshifted
is clearly evident.}
\label{polar}
\end{figure}

\begin{figure*}
\includegraphics[scale=0.45, angle=0]{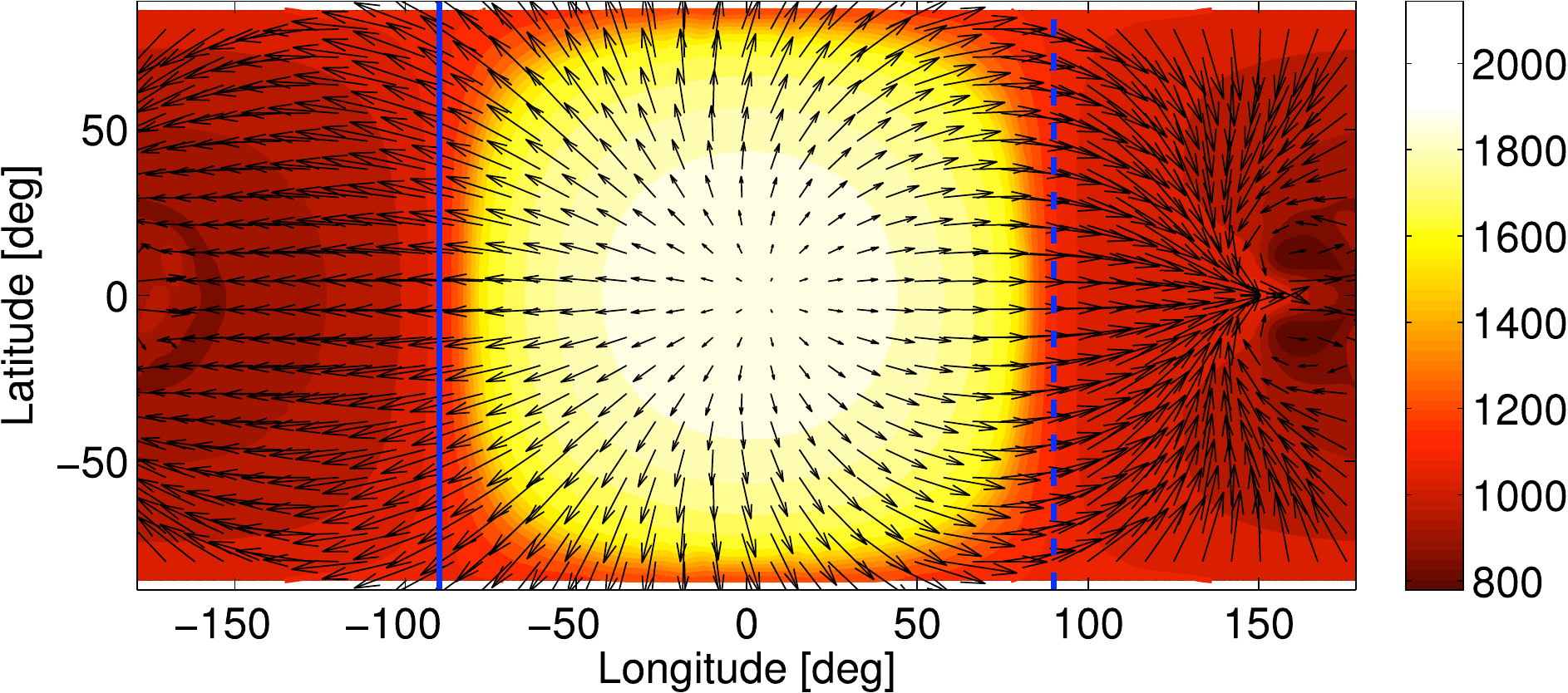}
\includegraphics[scale=0.45, angle=0]{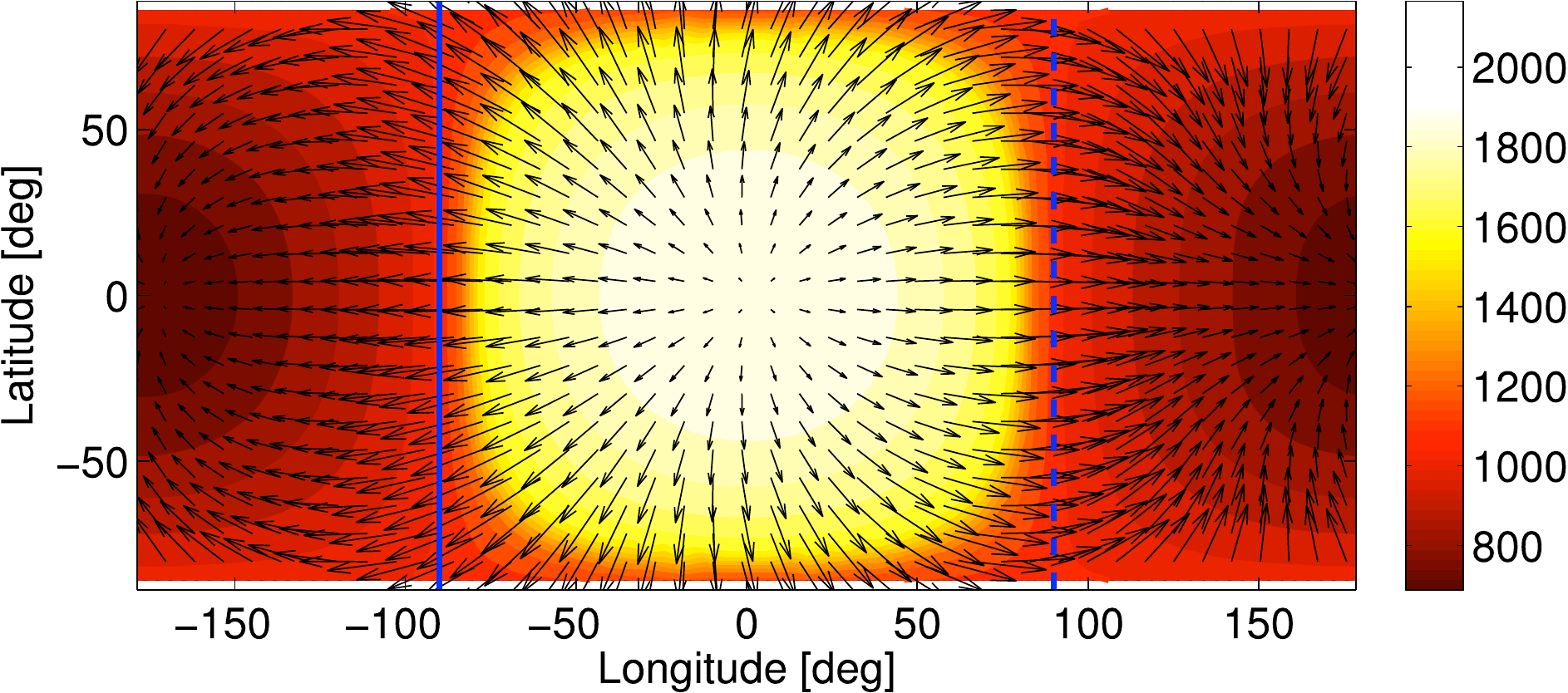}\\

\includegraphics[scale=0.45, angle=0]{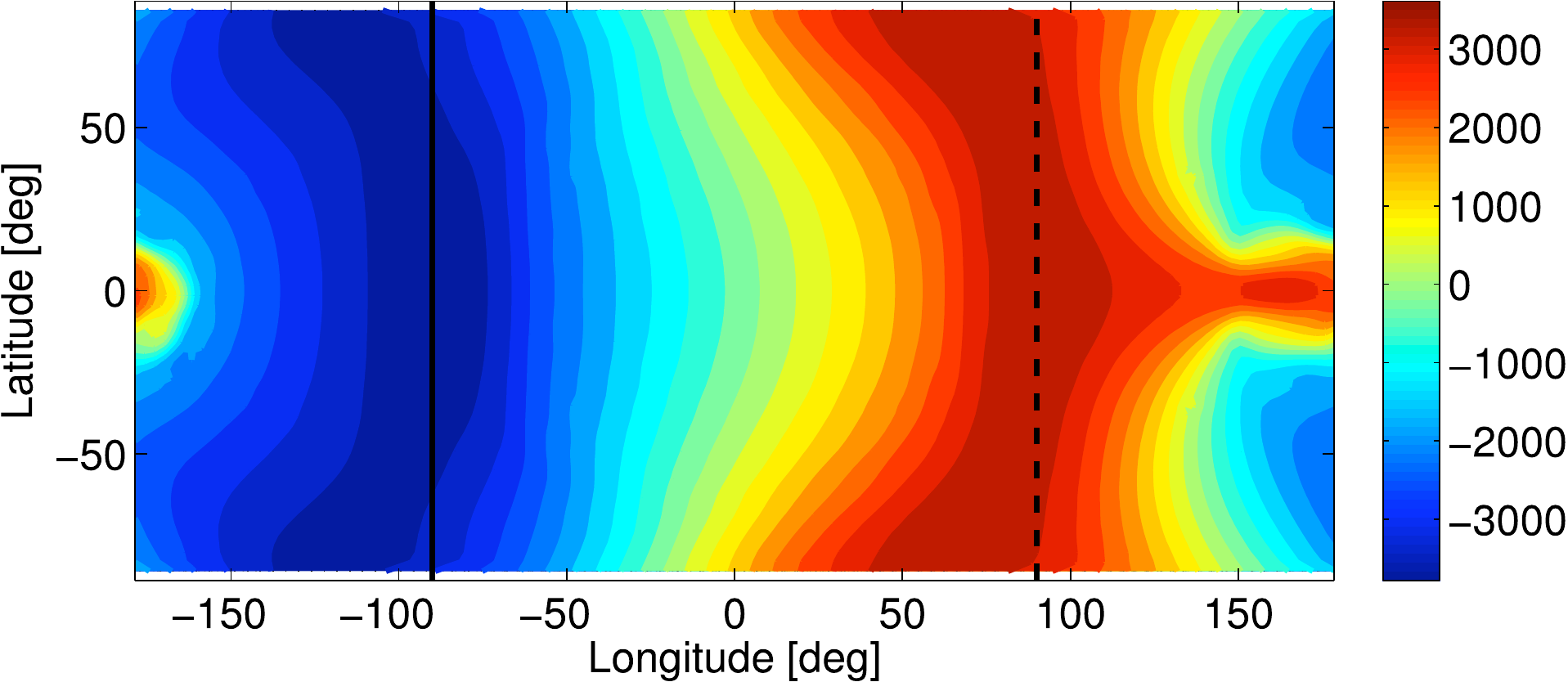}
\includegraphics[scale=0.45, angle=0]{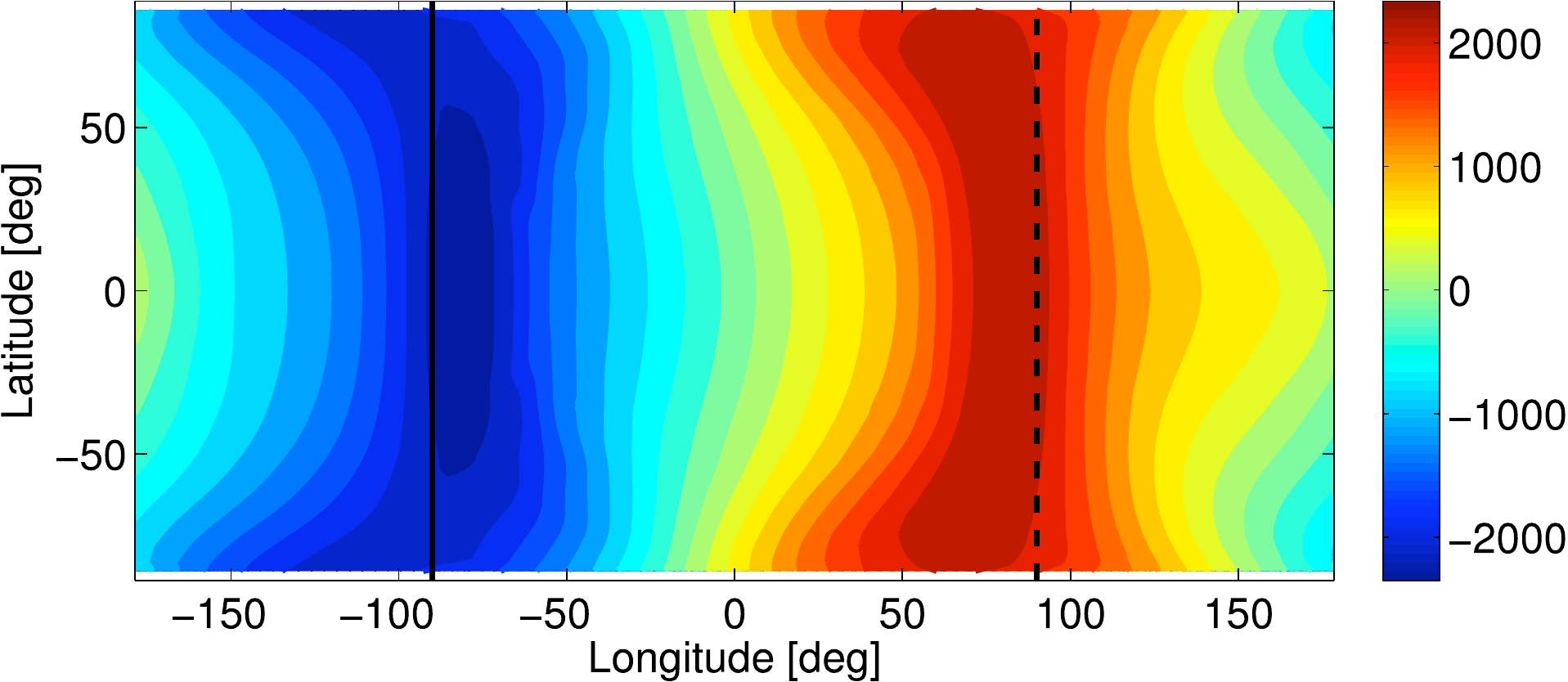}

\caption{Temperature and winds at 0.1 mbar pressure in models of HD 209458b 
with frictional drag.  Left: a model with a drag time constant of
$3\times10^4\,$s.  Right: a model with a drag time constant of
$10^4\,$s.   Top panels show temperature in K (orange scale) and winds
(arrows).  Bottom panels show zonal wind in $\rm m\,s^{-1}$.  In the
case with weaker drag, an equatorial jet extends partway across the
nightside, but the jet is damped out in the case with stronger drag.
Vertical solid and dashed lines show the terminators.}
\label{drag}
\end{figure*}

To summarize, these models exhibit a transition from a circulation
dominated by zonal jets at modest insolation (GJ 436b) to one
dominated by day-night flow at high insolation (HD 209458b).
Qualitatively, this transition matches well the predictions from our
theory in Section 2---as the stellar insolation increases, the
effective radiative timescale decreases, and this damps the standing
planetary-scale Rossby and Kelvin waves, limiting their ability to
drive a dominant zonal flow and leading to a circulation comprised
primarily of day-to-night flow at low pressure.  We emphasize that the
models in Figures~\ref{temp-winds}--\ref{histogram} do not
contain frictional drag at the low pressures sensed by remote 
measurements, and so the only source of damping
is the radiation (as well as the Shapiro filter, which exerts minimal
effect at large scales).  The models show that the regime transition occurs
very gradually as stellar insolation is varied
(Figures~\ref{temp-winds}a--d).  This is also
consistent with theoretical expectations; as shown in
Figure~\ref{jeteddy}, when large-scale drag is absent, the radiative
time constant must be decreased by over a factor of $\sim$30 (from
$\sim$3 days to less than 0.1 day) to force the flow from the
jet-dominated to eddy-dominated regime.
Moreover, as discussed in Section~\ref{theory}, damping through
radiation alone can inhibit differential zonal propagation of the
planetary-scale waves, but the multi-way
horizontal force balance between pressure-gradient, Coriolis, and advective
forces can still produce prograde phase tilts near the equator.  
Thus, we still expect a narrow equatorial jet over at least some
longitudes.  This can be seen in the nonlinear shallow-water solutions
(see Figure~\ref{stswm-no-drag}(d)) and also explains 
the continuing existence of a narrow equatorial jet
even for extreme radiative forcing in the 3D models 
(Figure~\ref{temp-winds}(d)).

This regime transition manifests clearly in plots of terminator winds.
Figure~\ref{polar} shows the wind component projected along the line
of sight to Earth at the terminator for a sequence of models.  Red
represents velocities away from Earth (hence redshifted) while blue
represents velocities toward Earth (hence blueshifted).  The first,
second, and third rows of Figure~\ref{polar} show our nominal models
of GJ 436b, HD 189733b, and HD 209458b, while the fourth row depicts
our model of HD 209458b adopting frictional drag with $\tau_{\rm
  drag}=10^4\rm\,s$.  For GJ 436b, the leading limb is redshifted
while the trailing limb is blueshifted.  For HD 189733b, the
redshifted portion---corresponding to the equatorial jet---is confined
to the low and midlatitudes on the leading limb. As a result, at high
altitudes, only about one-quarter of the limb is redshifted, while
about three-quarters is blueshifted.  For our nominal HD 209458b
model, the confinement of the equatorial jet to low latitudes on the
leading limb is even stronger, such that only about $\sim$10\% of the
high-altitude limb is redshifted while $\sim$90\% is blueshifted.  In
the HD 209458b model with frictional drag, the high-altitude winds are
blueshifted over the entire terminator, completing the transition from
a circulation dominated by jets to one dominated by high-altitude
day-to-night flow.

The regime transition discussed here is affected not only by the stellar insolation
but also the atmospheric metallicity.  Larger metallicities
imply larger gaseous opacities (due to the increased abundance of H$_2$O, CO,
and CH$_4$), and this moves the photosphere to lower pressures
\citep[cf][]{spiegel-etal-2010, lewis-etal-2010}, implying that
the bulk of the starlight is then absorbed in a region with very
little atmospheric mass.  As a result,
increasing the atmospheric metallicity enhances the 
dayside heating and nightside cooling per mass at the photosphere
even when the stellar insolation remains unchanged.
The effects of this are illustrated
in Figure~\ref{temp-winds}b, which shows a GJ 436b model identical
to that in Figure~\ref{temp-winds}a except that the metallicity 
is 50 times solar \citep{lewis-etal-2010}.  Because of the greater
absorption of stellar radiation at high levels, the atmosphere exhibits
a large day-night temperature difference and significant 
zonal-wavenumber-one structure in the zonal wind, with
strong longitudinal variations in the equatorial jet reminiscent
of that in our HD 189733b model (compare Figures~\ref{temp-winds}(b)
and (c)).  Although eastward flow still dominates along
most of the terminator, as in the solar-metallicity GJ 436b model, 
the western terminator exhibits westward flow within $\sim$$30^{\circ}$
latitude of the pole. Spectral lines as seen during transit still exhibit
bimodel blue and redshifts, but the blueshifts are now slightly more
dominant (Figure~\ref{histogram}b).

Although we have focused on the existence of a regime transition in models
with differing stellar fluxes, it is worth emphasizing that the
same transition often occurs {\it within} a given model from
low pressure to high pressure.  Generally speaking, the radiative
time constants are short at low pressure and long at high pressure
\citep{iro-etal-2005, showman-etal-2008a}.  The theory presented
here therefore predicts that, as long as the incident stellar flux
is sufficiently high and frictional drag is sufficiently weak, the air should
transition from a day-to-night flow pattern at low pressure to
a jet-dominated zonal flow at high pressures.  Just such a pattern
is seen in many published 3D hot Jupiter models \citep[e.g.,][]{cooper-showman-2005,
cooper-showman-2006, showman-etal-2008a, showman-etal-2009, rauscher-menou-2010,
heng-etal-2011}.  Note, however, that if the incident stellar flux is sufficiently low
(and the drag is very weak),
the atmosphere may be in a regime of jet-dominated flow throughout; on
the other hand, if frictional drag is sufficiently strong, jets may be unable to 
form at all,
and the atmosphere may be in a regime of day-night flow aloft with a very
weak return flow at depth. 

\subsection{Results: influence of drag}
\label{results-drag}

We now consider the effect of frictional drag in 3D models.   As discussed in
Section~\ref{theory}, sufficiently strong frictional drag  (i) 
damps the standing planetary-scale waves that are the natural response
to the day-night
heating gradient,  and (ii) drives the horizontal force balance into a two-way balance between pressure-gradient
and drag forces, both of which inhibit the development of prograde phase
tilts in the eddy velocities and in turn the pumping of zonal jets,
and, finally (iii) directly damps the zonal jets.
Thus, we expect that an atmosphere with sufficiently strong frictional
drag will lack zonal jets and that its circulation will instead
consist primarily of day-to-night flow at high altitude, with return
flows at depth.  Figure~\ref{drag} illustrates this for
solar-metallicity models of HD 209458b where Rayleigh drag is implemented
with time constants of $3\times 10^4\rm\,s$ (left column) and
$10^4\rm\,s$ (right column).  
As predicted, the air at 0.1 mbar flows directly from dayside to
nightside over both terminators.  The model with $\tau_{\rm drag}=3\times10^4\rm\,s$ exhibits
a remnant equatorial jet on the nightside that extends from the eastern terminator
to the antistellar point.  Because the stronger frictional drag damps
it out, the model with $\tau_{\rm drag}=10^4\rm\,s$ lacks such a jet,
and the flow exhibits only modest asymmetry (due to the $\beta$
effect) between the western and eastern terminators.  Doppler 
lines would be entirely blueshifted in both cases.

\begin{figure}
\includegraphics[scale=0.6, angle=0]{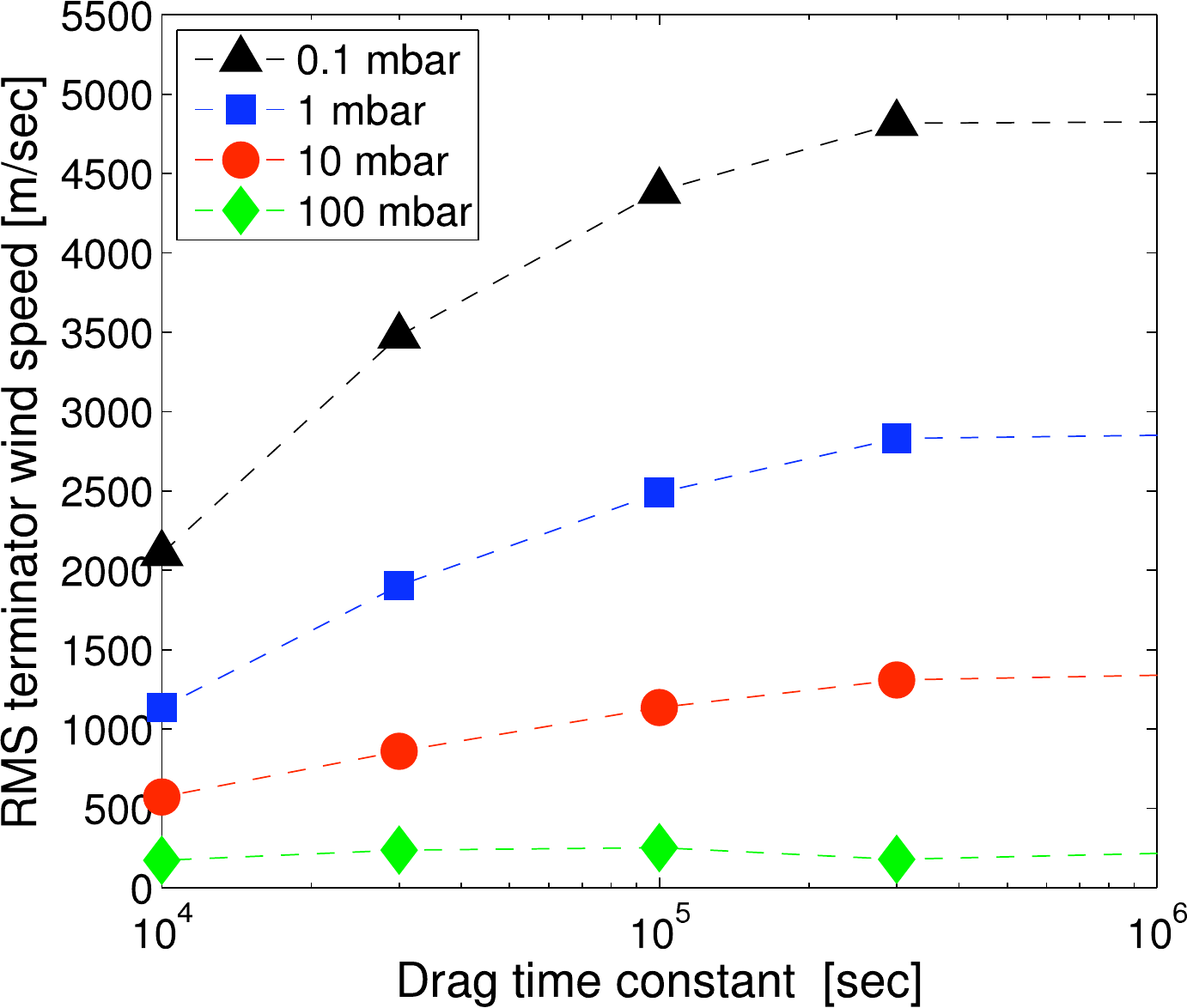}

\caption{Steady-state root-mean-square wind speeds at the terminator
  versus frictional drag time constant from a sequence of HD 209458b
  models including drag.  For each 3D model, performed for a given
  drag time constant, the root-mean-square wind speeds---calculated
  along the terminator---are shown at pressures of 0.1, 1, 10, and 100
  mbar.  For these models the speeds generally represent day-night
  flow.  The dashed lines to the right of the rightmost points are
  connecting to the model with no drag in the upper atmosphere
  ($\tau_{\rm drag}\to\infty$), where the rms terminator wind speed is 5200,
  3800, 2600, and $1900\rm \,m\,s^{-1}$ at 0.1, 1, 10, and 100 mbar,
  respectively.  The equilibrated speeds depend significantly on the
  drag time constant and, within a given model, on pressure.  }
\label{drag2}
\end{figure}

Friction affects not only the qualitative circulation regime (e.g.,
existence or lack of zonal jets) but also the speed of the
high-altitude flow between day and night.  Figure~\ref{drag2} shows
the root-mean-square zonal wind speeds at the terminator for a
sequence of HD 209458b models with differing drag time constants.  All
of these runs are in the same regime as the model in
Figure~\ref{drag}, where day-to-night flow dominates at low pressure.
When $\tau_{\rm drag}$ is sufficiently long, the flow speeds are
independent of the drag time constant, but they start to decrease when
$\tau_{\rm drag}$ is sufficiently short.  When drag is absent in the
upper atmosphere, our HD 209458b model equilibrates to an rms terminator wind
speed of $5.2\rm\,km\,s^{-1}$ at 0.1 mbar, decreasing with depth
to $3.8$, $2.6$, and $1.9\rm\,km\,s^{-1}$ at 1, 10, and 100 mbar,
respectively.  As shown in Figure~\ref{drag2}, the addition of 
weak drag ($\tau_{\rm drag}=3\times10^5\rm\,s$)
exerts only a modest effect on the day-night flow speeds at pressures
$\lesssim 10\rm\,mbar$. Drag time constants $\tau_{\rm drag}\lesssim 10^5\rm\,s$
start to matter significantly in the upper atmosphere, however;
for example, for $\tau_{\rm drag}=10^4\rm\,s$, the rms terminator
speeds are $2.1$, $1.1$, $0.6$, and $0.2\rm\,km\,s^{-1}$ at 0.1, 1,
10, and 100 mbar---significantly less than the equilibrated speeds
in the absence of upper-level drag.

The above results suggest that the amplitude of the observed Doppler
shift can place constraints on the strength of frictional drag in the
upper atmospheres of hot Jupiters.  \citet{snellen-etal-2010}'s
inference of winds on HD 209458b is tentative but, at face value,
suggests wind speeds toward Earth of $2\pm 1\rm\,km\,s^{-1}$.
\citet{snellen-etal-2010} suggest that their measurements are sensing
pressure levels of 0.01--0.1 mbar.  At these levels, the winds in our
models equilibrate to $\sim$4--$6\rm\,km\,s^{-1}$ when drag is weak,
and Figure~\ref{drag2} shows that reducing the wind speed to
$2\rm\,km\,s^{-1}$ requires drag time constants potentially as short
as $\sim$$10^4\rm\,s$.  This hints that strong frictional drag processes
may operate in the atmosphere of HD 209458b.
But caution is warranted:  Figure~\ref{drag2} also
demonstrates that the rms terminator wind speeds also depend strongly
on pressure within any given model; therefore, making robust
inferences about drag amplitudes from observed Doppler shifts
requires extremely careful and
accurate estimates of the pressure levels being probed.  This may
be a challenge, at least until the composition and hence wavelength-dependent
opacity of hot Jupiters are better understood.  
If the \citet{snellen-etal-2010} measurements
are actually sensing deeper pressures of $\sim$$10\rm\,mbar$,
say, then explaining their 2-$\rm km\,s^{-1}$ signal would
require little if any drag in the observable atmosphere.  

In light of Figure~\ref{drag2}, it is interesting to briefly comment
on the pressures being probed in transmission spectra computed from our
models.  In Section~\ref{observables}, we will present transmission
spectra for our 3D models computed self-consistently from high-spectral-resolution
versions of the same
opacities used to integrate the GCM.  These calculations indicate that,
in the $K$-band region considered by Snellen et al.,
our synthetic transmission spectra probe pressures ranging from $\sim$10 mbar
in the continuum between spectral lines to less than $\sim$0.1 mbar at line
cores.  It is the Doppler shifts of the spectral lines that
are observable---the Doppler shift of the continuum, if any, is almost
undetectable since the absorption depends only weakly on wavelength
there.  As a result, the overall Doppler signal detected in a spectral cross-correlation
is heavily weighted toward the
Doppler shift of the spectral lines.  We find that, when cross-correlating
our synthetic transmission spectra with template spectra, our models of 
HD 209458b primarily probe the atmospheric winds at pressures of 0.1 to 1 mbar.

The qualitative dependence of terminator wind speed on the drag time
constant---illustrated in Figure~\ref{drag2}---can be understood
analytically. To order of magnitude, the horizontal pressure gradient
force in pressure coordinates between day and night can be written
$R\Delta T_{\rm horiz}\Delta \ln p/a$.  This is balanced by some
combination of advection, of magnitude $U^2/a$, Coriolis force, of
magnitude $fU$, and drag, of magnitude $U/\tau_{\rm drag}$.  Drag will
dominate when $\tau_{\rm drag}\lesssim f^{-1}$ and when $U/\tau_{\rm
  drag}\gtrsim U^2/a$, which requires $\tau_{\rm drag} \ll
(a/|\nabla\Phi|)^{1/2}$, equivalent to the requirement that $\tau_{\rm
  drag} \ll a/(R\Delta T_{\rm horiz} \Delta \ln p)^{1/2}$.  As long as
these conditions are satisfied, we can balance drag against the
pressure-gradient force.  Solving for $\tau_{\rm drag}$ then implies
that the amplitude of drag necessary to obtain a wind speed $U$ is
\begin{equation}
\tau_{\rm drag} \sim {Ua\over R\Delta T_{\rm horiz}\Delta \ln p}
\label{drag-strength}
\end{equation}
Inserting parameters appropriate to the 0.1 mbar level on HD 209458b
($a\sim10^8\rm\,m$, $R\sim3700\rm\, J\,kg^{-1}\,K^{-1}$, $\Delta
T_{\rm horiz}\sim 1000\rm\,K$, and $\Delta \ln p\sim 5$), and adopting
$U\sim 2\rm\,km\,s^{-1}$ motivated by the Snellen et al. measurement
of HD 209458b, we obtain $\tau_{\rm drag}\sim 10^4\rm\,s$.  This value
agrees well with the strength of drag needed in our 3D model
intregrations to achieve a speed of $2\rm\,km\,s^{-1}$ at the 0.1 mbar level
(leftmost black triangle in Figure~\ref{drag2}).

\section{Transmission Spectrum Calculations}

To quantify the implications for observations, in this section
we present theoretical transmission spectra from our 3D models
demonstrating the influence of Doppler shifts due to atmospheric winds.
These spectra illustrate how the dynamical regime shifts described in the
preceding sections manifest in transit spectra.

\label{observables}
\subsection{Methods}

We have previously developed a code to compute the transmission
spectrum of transiting planet atmospheres, which we extend here
to include Doppler shifts due to atmospheric winds. The first generation of
the code, which used one-dimensional atmospheric pressure-temperature
(\emph{p-T}) profiles, is described in \citet{hubbard-etal-2001} and
\citet{fortney-etal-2003}.  In \citet{shabram-etal-2011} the one-dimensional
(1D) version of the code was well-validated against the analytic
transmission atmosphere model of \citet{lecavelier-des-etangs-etal-2008}.  In
\citet{fortney-etal-2010} we implemented a method to calculate the transmission
spectrum of fully 3D models.

The calculation of the absorption of light passing through the
planet's atmosphere is based on a simple physical picture.  One can
imagine a straight path through the planet's atmosphere, parallel to
the star-planet-observer axis, at an impact parameter $r$ from this
axis.  The gaseous optical depth $\tau_{\rm G}$, starting at the
terminator and moving outward in one direction along this path, can be
calculated via the equation:
\begin{equation} 
\label{taum}
\tau_{\rm G}=\int_{r}^{\infty}\frac{r^{\prime}
\sigma(r^{\prime}) n(r^{\prime})}
{{(r^{\prime2}-r^{2})}^{1/2}} \,dr^{\prime},
\end{equation}
where $r^{\prime}$ is the distance between the local location in the
atmosphere and the planetary center, $n$ is the local number density
of molecules in the atmosphere, and $\sigma$ is the
wavelength-dependent cross-section per molecule.  Later we will
discuss the role of winds leading to a Doppler shifted $\sigma$ away
from rest wavelengths.  We assume hydrostatic equilibrium with a
gravitational acceleration that falls off with the inverse of the
distance squared.  The base radius is taken at a pressure of 10 bars, where the
atmosphere is opaque, and this radius level is adjusted to yield the
best fit to observations, where applicable.  Here we define the
wavelength-dependent transit radius as the radius where the total
slant optical depth reaches 0.56, following
\citet{lecavelier-des-etangs-etal-2008}.  Additional detail and
description can be found in \citet{fortney-etal-2010}, as the 3D setup
here is the same as described in that paper.

For any particular column of atmosphere, hydrostatic equilibrium is
assumed, and we use the given local \emph{p-T} profile to interpolate
in a pre-tabulated chemical equilibrium and opacity grid that extends
out to 1 $\mu$bar.  The equilibrium chemistry mixing ratios
\citep{lodders-1999, lodders-fegley-2002,lodders-fegley-2006} are
paired with the opacity database \citep{freedman-etal-2008} to
generate pressure~-~, temperature-, and wavelength-dependent
absorption cross-sections that are used for that particular column.
For a different column of atmosphere, with a different \emph{p-T}
profile, local pressures and temperatures will yield different mixing
ratios and wavelength-dependent cross-sections.

We include the Doppler shifts due to the local atmospheric winds and
planetary rotation when evaluating the opacity at any given region of
the 3D grid.  At high spectral resolution, rotation tends to cause a
broadening of spectral lines \citep{spiegel-etal-2007}, while the
atmospheric wind speeds lead to absorption features that are Doppler
shifted from their rest wavelengths
\citep{snellen-etal-2010,kempton-rauscher-2012}.  The cross section
$\sigma$ is not evaluated at the rest wavelength, $\lambda_0$, but
rather at the Doppler shifted value, $\lambda$, found via
\begin{equation} \label{shift}
\lambda=\lambda_0 (1-\frac{v_{\rm los}}{c}),
\end{equation}
where $v_{\rm los}$ is the line-of-sight velocity---including both
rotation and atmospheric winds---and $c$ is the speed of light.  The
\citet{snellen-etal-2010} observations were performed at a resolving
power of $R\sim10^5$.  For additional clarity in presentation, we have
computed opacities and transmission spectra at $R = 10^6$.  In
practice we interpolate within our $R=10^6$ opacity database to yield
the correct $\sigma$ for every height in the atmosphere, on every
column, given the calculated velocities at every location in our grid.
This is done at 128 locations around the terminator.  The contribution
to the transmission spectrum is strongly weighted toward regions near
the terminator, and falls essentially to zero more than
$\sim$$20^{\circ}$ from the terminator (where the transit chord
reaches extremely low pressures).  Therefore, we only include in the
calculation regions within $\pm20^{\circ}$ of the terminator (i.e., a
total swath $40^{\circ}$ wide centered on the terminator).  Note that
for simplicity we do not include the Doppler shift due to orbital
motion, and we are therefore essentially evaluating the transmission
spectrum at the center of the transit for a planet with zero orbital
eccentricity.  The effect of orbital motion was considered by
\citet{kempton-rauscher-2012}.

\subsection{HD 209458b Cases and the Role of Drag}
\begin{figure*}
\plotone{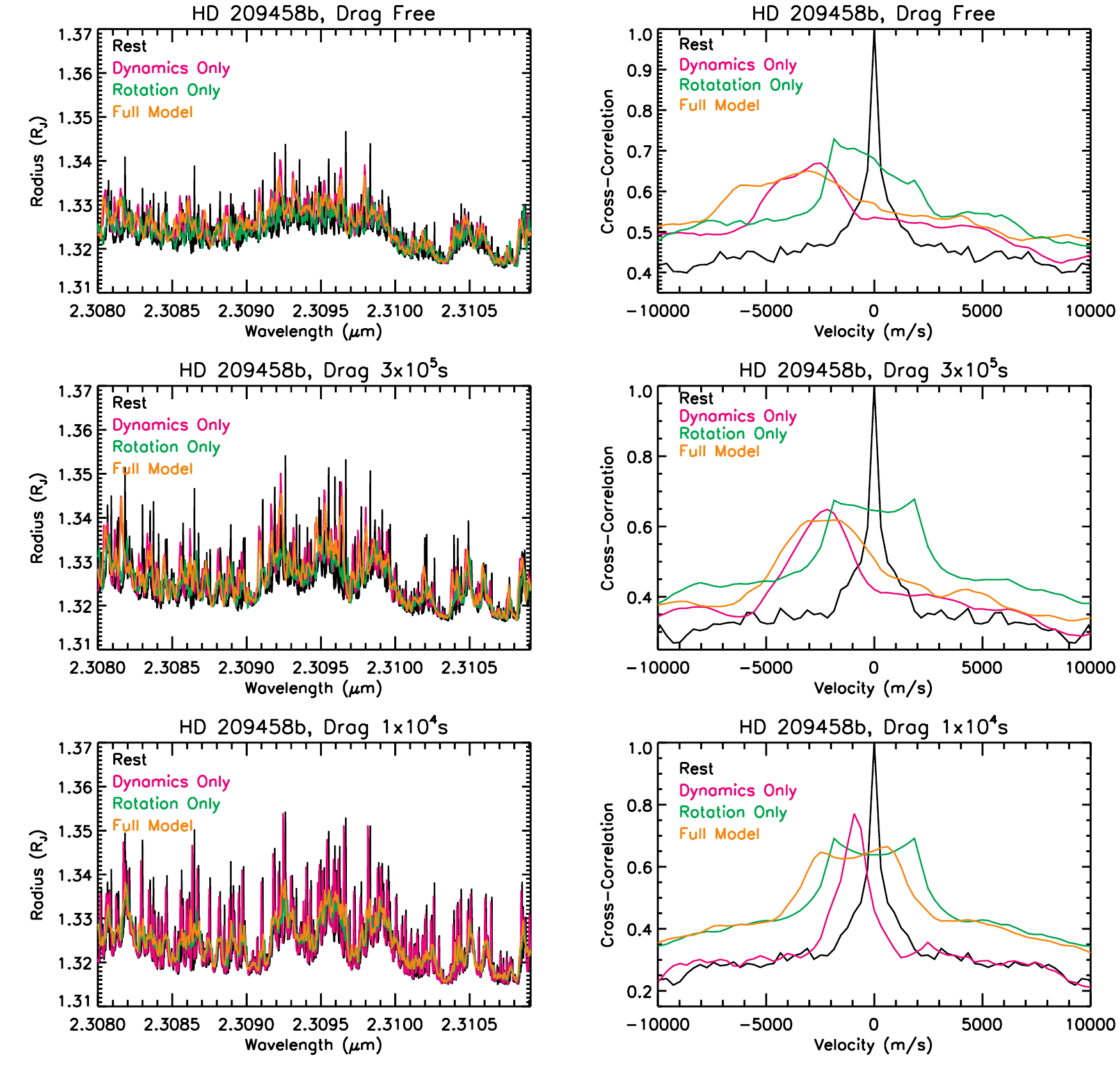}
\caption{\emph{Left panels}: High spectral resolution ($R\sim~10^6$)
  planet radius versus wavelength for three models of HD 209458b with
  various drag strengths.  Solid black is the ``Rest'' model
  calculated without a Doppler shift.  Magenta includes Doppler shifts
  due to atmospheric dynamics only.  Green includes Doppler shifts due
  to planetary rotation only.  In orange, the ``Full Model" includes
  both dynamics and rotation.  \emph{Right panels}: Cross-correlations
  for the planetary transmission spectra shown at left.  As drag
  becomes stronger one generally finds slower and more peaked wind
  speeds.  See text for additional details.
\label{209cases}}
\end{figure*}

We now turn to a detailed analysis of the HD 209458b models in the
vicinity of the \citet{snellen-etal-2010} observations of HD 209458b
near 2.2 $\mu$m.  The Doppler shift was not measured across all
wavelengths, but only within the narrow CO lines, since flat
transmission spectra (corresponding to the continuum between the
spectral lines) yield no leverage on the Doppler shifts.  These peaks
are all of nearly the same strength \citep[see, e.g.][supplemental
  online material]{snellen-etal-2010}, so they probe very similar
heights in the atmosphere.  In Figure \ref{209cases} we have computed
the transmitted spectrum at $R=10^6$, 1300 wavelengths, from 2.3080 to
2.3011 $\mu$m, for three models of HD 209458b: a model with no drag in
the upper atmosphere (top left panel), a weak-drag model with a drag
time constant of $3 \times 10^5\rm\,s$ (middle left panel), and a
strong-drag model with a drag time constant of $1 \times 10^4\rm\,s$
(bottom left panel).

For each of the three drag cases we calculated the model planet radius
vs.~wavelength for four variants of the dynamical model.  One uses
rest-wavelength values (black, this corresponds to a reference case where winds
are assumed to be zero), one includes atmospheric dynamics only
(magenta, ignoring rotation but using the full 3D winds), one includes 
\emph{only} rotation (green, ignoring dynamics), and in orange is the 
full model, including both dynamics and rotation.  The transmission spectra in Figure
\ref{209cases} are somewhat difficult to interpret.  Therefore we have
also calculated the cross-correlation, compared to the rest wavelength
model, across the 1300 wavelengths in our calculation.

There are several aspects of note for these plots.  Starting in the
upper right of Figure~\ref{209cases}, the drag free HD 209458b model,
we see that the self cross-correlation is strongly peaked at 0 m
s$^{-1}$, as expected.  The dynamics-only model shows winds that
peaked at $-2500\rm\, m\,s^{-1}$ (meaning a blueshift), with strong
winds (generally at the equator) reaching beyond $-5000\rm\,
m\,s^{-1}$.  As seen in the dynamics output, there is also a component
of redshift winds, taking up a relatively small fraction of the
terminator, which range from $\sim$0 to $5000\rm\,m\,s^{-1}$.
Interestingly, for this drag-free case, the cross-correlation curve of
the rotation-only model is not symmetric about the zero-velocity
point.  This asymmetry only appears in models with a strong
leading/trailing hemispheric temperature contrast.  It appears to be
due the trailing (hotter) hemisphere having a larger scale height, and
therefore more prominent absorption features.  The full model,
including dynamics and rotation, has a broader peak than the
dynamics-only model due to rotational broadening.  The peak
$-5000\rm\, m\,s^{-1}$ velocities from dynamics and $-2000\rm\,
m\,s^{-1}$, from rotation, lead to velocities on the trailing
hemisphere's equator of $-7000\rm\, m\,s^{-1}$.  The full model is not
merely just a broadened dynamics-only model due to the asymmetric
rotational component.

The weak-drag case (middle right of Figure~\ref{209cases}) has a more
constrained atmospheric flow, which is generally day-to-night, with a
much reduced super-rotating jet.  Velocities from dynamics are peaked
more narrowly around $-2200\rm\, m\,s^{-1}$ (red curve).  The
rotational component is close to symmetric, due to the small
leading/trailing temperature difference.  The full model looks much
like the dynamics-only model, broadened due to planetary rotation.

\begin{figure*}
\plotone{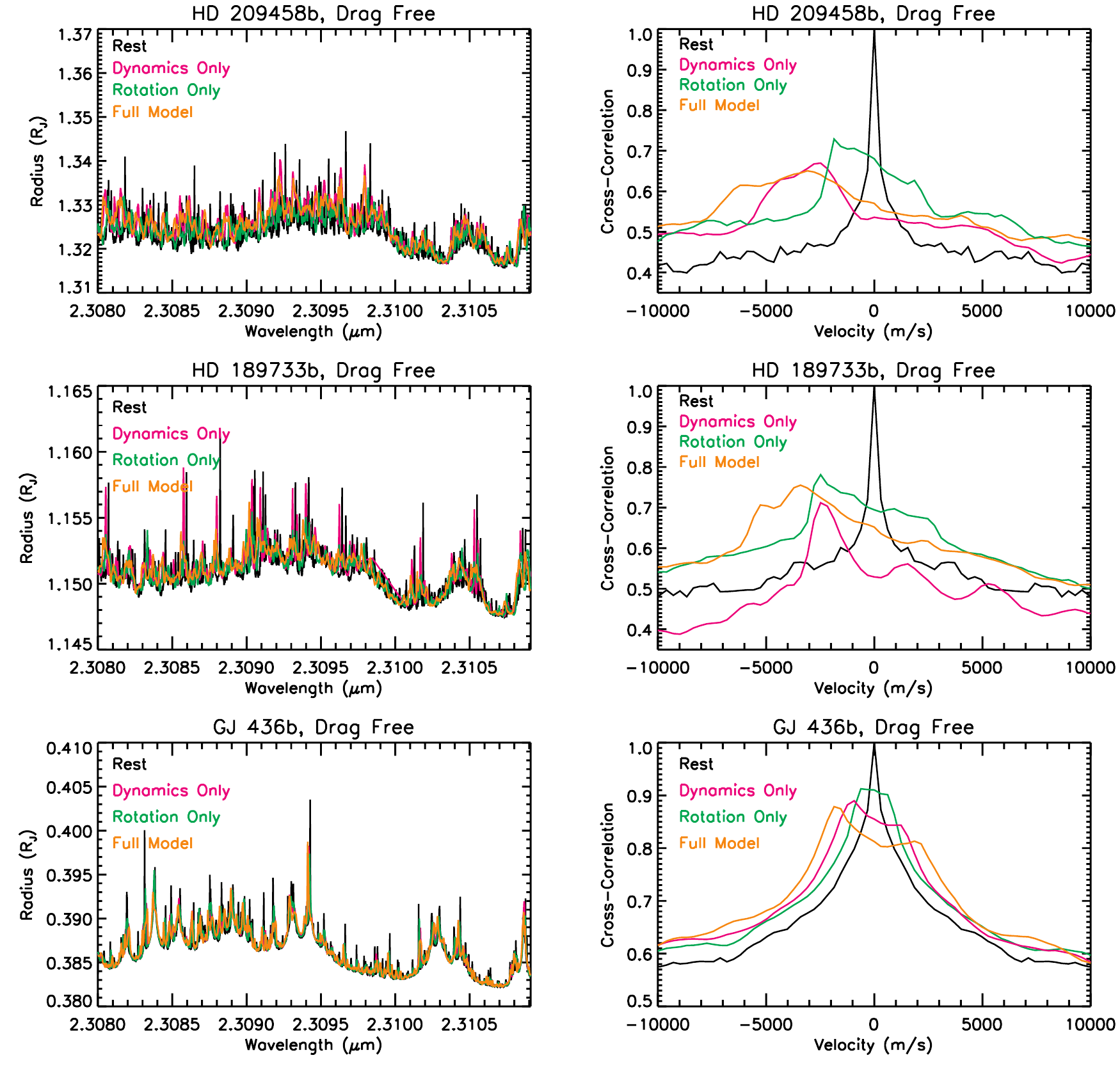}
\caption{\emph{Left panels}: High spectral resolution ($R\sim~10^6$)
  planet radius versus wavelength for drag-free models of HD 209458b,
  HD 189733b, and GJ 436b.  Solid black is the ``Rest'' model
  calculated without a Doppler shift.  Magenta includes Doppler shifts
  due to atmospheric dynamics only.  Green includes Doppler shifts due
  to planetary rotation only.  In orange, the ``Full Model" includes
  both dynamics and rotation.  \emph{Right panels}: Cross-correlations
  for the planetary transmission spectra shown at left.  As the
  incident flux becomes smaller, from HD 209458b, to HD 189733b, to GJ
  436b, peak wind speeds become smaller, and the planet's dynamics
  become more dominated by a super-rotating jet.  See text for
  additional details.
\label{3cases}}
\end{figure*}

The strong-drag case (lower right of Figure~\ref{209cases}) has a very
constrained circulation, with a relatively uniform day-to-night flow
around the entire planet, with little sign of an equatorial jet.  The
dynamics velocities are strongly peaked at $-1000\rm\, m\,s^{-1}$ and
the small leading/trailing temperature contrast leads to a
symmetric rotational component.  The full model looks very much like a
broadened version of the dynamics-only model.  The slightly higher
peak on the right side of the full model is due to a slight asymmetry
in the velocities around $-1000\rm\,m\,s^{-1}$ in the dynamics-only
model.

Overall, a comparison of the models in Figure~\ref{209cases}
highlights the possibility that the amplitude of drag in the
atmosphere of HD 209458b can be inferred from observations.  The
no-drag, weak-drag, and strong-drag models (Figure~\ref{209cases})
exhibit peaks in the cross-correlation centered at $-4$, $-2.5$, and
$-1\rm\,km\,s^{-1}$, respectively.  We emphasize that these values are
the quantitative result of our rigorously calculated transmission
spectra from our fully coupled 3D models (and are not, for example,
simply the velocity at some assumed pressure of the 3D models).  At
face value, the results in Figure~\ref{209cases} indicate that models
with a drag time constant of $10^4$--$10^6\rm\,s$ provide a better fit
to the \citet{snellen-etal-2010} observations than models with no drag
in the upper atmosphere.  This is consistent with the inferences drawn
in Section~\ref{results-drag}.

The possibility of drag in the atmosphere of HD 209458b is particularly
interesting in light of recent suggestions that thermal ionization
of alkali metals at high temperature can lead to Lorentz forces that
act to brake the atmospheric winds \citep{perna-etal-2010, menou-2012}.
In the regime of day-to-night flow, air at the terminator has just crossed
much of the dayside, and so the wind speed at the terminator 
is predominantly determined by drag on the dayside rather than the
nightside.   Secondary-eclipse observations
indicate that HD 209458b exhibits a dayside stratosphere with temperatures
potentially reaching $\sim$$2000\rm\,K$ (\citet{knutson-etal-2008};
see also our Figure~\ref{temp-winds}).  Based on the scaling relations
in \citet{perna-etal-2010}, such high temperatures should lead to very
short drag times, potentially consistent with the inferences on 
drag drawn here.

\subsection{Comparing Three Different Planets}
Figure \ref{3cases} allows us to diagnose the different atmospheric
dynamics and Doppler shift signatures of HD 209458b, HD 189733b, and
GJ 436b.  All cases are drag-free.  The top row is the same HD 209458b
model described in the top row of Figure \ref{209cases}.  The
dynamical wind velocities for HD 189733b (middle panels of Figure~\ref{3cases}) 
are relatively similar to those of HD 209458b, but lacking a
very high velocity component.  The planet's rotation period of 2.2
days is only 63\% of the period of HD 209458b, meaning HD 189733b has
a significantly larger rotational velocity component.  The asymmetry
shown is also due to a relatively large leading/trailing temperature
contrast.  The full model has a smaller, high velocity peak at -5000 m
s$^{-1}$ due to the strong blue-shifted peak of rotational velocity.

The GJ 436b spectrum clearly shows a transition to a different regime
of atmospheric dynamics.  The flow is dominated by a wide
super-rotating jet, with little flow purely being day to night.  This
manifests itself in the dynamics-only model as being somewhat
symmetric, with a slightly higher cross-correlation peak on the
blue-shifted side.  The rotational component is nearly symmetric,
owning to relative temperature homogenization of the planet.  The
magnitude of the rotational velocities are quite small because the
planet has a small radius.  The full model shows little Doppler shift.
The spectrum plot at left shows very little difference between all of
the models, other than that they have been Doppler broadened compared
to the rest model.

\section{Discussion}
\label{discussion}

Recent observations suggest that atmospheric winds on hot Jupiters can
be directly inferred via the Doppler shift of spectral lines seen
during transit \citep{snellen-etal-2010}.  Motivated by these
observations, we have shown that the atmospheric circulation of hot
Jupiters divides into two regimes depending on the strength of the
radiative forcing and frictional drag, with implications for the
Doppler signature:
\begin{itemize}

\item Under moderate stellar fluxes and weak to moderate drag,
  atmospheric waves generated by the day-night thermal forcing
  interact with the mean flow to produce fast east-west (zonal)
  jets. In this regime, air along the terminator flows toward Earth in
  some regions and away from Earth in others, leading to blue-shifted
  and red-shifted contributions to the Doppler signature seen during
  transit.  Depending on the speed of the winds relative to the
  planetary rotation, as well as the variation of zonal winds in
  latitude and height, this will cause Doppler lines observed during
  transit to be broadened or, in extreme cases, split into distinct,
  superposed blue-shifted and red-shifted velocity peaks.

\item Under extreme stellar fluxes and/or strong frictional drag,
  however, the radiative and/or frictional damping is so strong that
  it damps these waves and inhibits jet formation.  At the low
  pressures sensed by transit measurements, the atmospheric
  circulation then involves a day-to-night flow, with a return flow at
  deeper levels.  In this regime, the airflow at levels sensed by
  transit measurements is toward Earth along most or all of the
  terminator, leading to a predominantly blue-shifted Doppler
  signature of spectral lines observed during transit.
\end{itemize}
We presented a theory predicting this regime transition,
and we confirmed its existence and explored its properties in one-layer
shallow-water models and in three-dimensional models coupling the
dynamics to realistic non-gray radiative transfer.  We then
presented detailed radiative transfer calculations of the transit
spectra expected from our 3D models in the 2-$\mu$m spectral region
observed by \citet{snellen-etal-2010}; these calculations can help to guide
future observational efforts.

We also showed that, in the second regime described above, the speed
of the day-night windflow depends on the amplitude of the drag at the
low pressures sensed by transit measurements.  Under relatively weak
drag, the wind speeds at the terminator of our HD 209458b models reach
$\sim$4--$6\rm\,km\,s^{-1}$ depending on altitude and forcing
conditions.  Under strong drag, the wind speeds are slower.
Interestingly, at the low pressures sensed by transit observations,
the drag must be relatively strong---with effective drag time
constants of $\sim$$10^6\rm\,s$ or less---to reduce the speeds by a
significant fraction.  Our models of HD 209458b without significant
drag in the upper atmosphere produce peak cross-correlations of the
transit spectrum corresponding to blueshifts of
$\sim$3--$7\rm\,km\,s^{-1}$; this exceeds, albeit marginally, the
$\sim$$2\rm\,km\,s^{-1}$ blueshift inferred by
\citet{snellen-etal-2010}.  On the other hand, our models that agree
best with the \citet{snellen-etal-2010} inference---where the peak
cross-correlations of the transit spectrum lie at
$\sim$1--$3\rm\,km\,s^{-1}$---exhibit drag timescales of
$10^4$--$10^6\rm\,s$.  This suggests, tentatively, that frictional
drag may be important on the dayside of HD 209458b.  An attractive
possibility is that the dayside of HD 209458b is sufficiently hot for
partial ionization to occur, leading to Lorentz-force braking of the
winds \citep{perna-etal-2010, menou-2012}. Regardless, these models
demonstrate that, in principle, measurements of the Doppler shift of
spectral lines can place constraints on the amplitude of drag in the
atmospheres of hot Jupiters.

In light of this issue, we note that our results differ from those of
\citet{kempton-rauscher-2012}, who obtained peak cross-correlations
in the transmission spectra corresponding to blueshifted velocities of
about $-2.5\rm\,km\,s^{-1}$ and $-1\rm\,km\,s^{-1}$ in models without
and with drag, respectively.  Their velocity shifts for their drag-free case are
significantly slower than the velocity shifts we obtain of $-4\rm\,km\,s^{-1}$
for our drag-free HD 209458b model.  A significant difference in the
two studies is that, in \citet{kempton-rauscher-2012}, the
heating/cooling in the thermodynamic energy equation was determined
using a simplified Newtonian relaxation scheme based on that presented
in \citet{cooper-showman-2005}; in contrast, our dynamical models are
fully coupled to non-gray radiative transfer, from which the radiative
heating/cooling rates are calculated.  This may lead to a quantitative
difference in the radiative heating rates and hence three-dimensional
wind structure.  Future work may shed light on the discrepancies
between the results.

It is worth mentioning that, within each of the two broad dynamical regimes
studied in this paper, there may lie additional subregimes involving
important transitions between dynamical mechanisms.  For example, in
the regime of zonal jets, we have emphasized the development of
equatorial superrotation by standing, planetary-scale Rossby and
Kelvin waves induced by the day-night thermal forcing
\citep{showman-polvani-2011}.  However, when the stellar flux is lower
than considered in most hot-Jupiter models, the importance of the
day-night thermal forcing decreases and the equator-to-pole heating
gradient becomes dominant.  Baroclinic instabilities can then occur,
particularly when the rotation rate is fast, and these may lead to
multiple mid-latitude east-west jets, with an equatorial jet that may
be of either sign depending on the details.  A transition analogous to
this is evident in models of GJ 436b presented by
\citet{lewis-etal-2010}. We will explore such dynamical transitions
further in future work.

It is also worth discussing the proposal of
\citet{montalto-etal-2011}, who suggested that the
$\sim$$2\rm\,km\,s^{-1}$ blueshift inferred by
\citet{snellen-etal-2010} results not from atmospheric winds but from
planetary orbital motion due to an eccentric orbit.  The radial
velocity at the time the planet crosses the line of sight to Earth is
${\rm RV_0}=\tilde K e\cos\omega/\sqrt{1-e^2}$ where $e$ is the
eccentricity, $\omega$ is the argument of periastron, and $\tilde
K\equiv [2\pi G/P]^{1/3}M_\star\sin i / (M_\star+
m_p)^{2/3}=1.47\times 10^5\,\rm m\sec^{-1}$ is a constant, which we
have evaluated for HD 209458 parameters. Here, $P$ is the orbital
period, $G$ is the gravitational constant, $i$ is the orbital
inclination, and $M_\star$ and $m_p$ are the mass of the star and
planet, respectively. Explaining a 2-$\rm km\,\sec^{-1}$ blueshift
would thus require that $e\cos\omega\approx 0.014$.  \citet{montalto-etal-2011}
point out that the eccentricity itself is rather poorly constrained;
however, what matters is not eccentricity alone but the combination
$e\cos\omega$.  A key point, apparently not appreciated by \citet{montalto-etal-2011},
is that $e\cos\omega$ is
tightly constrained by observations of transit and secondary eclipse.
Observations of the relative timing of transit and secondary eclipse
from \citet{deming-etal-2005a} show that $e\cos\omega < 0.002$ at
1-$\sigma$.  Observations from \citet{knutson-etal-2008} and
\citet{crossfield-etal-2012} place even tighter upper limits on
$e\cos\omega$; the latter study yields $e\cos\omega =
0.00004\pm0.00033$, corresponding to a 3-$\sigma$ upper limit of the
orbit-induced Doppler shift of $140\rm\,m\,s^{-1}$ at the center of
transit.  This appears to rule out any orbital explanation for the
Doppler shift inferred by \citet{snellen-etal-2010}.

One also might wonder whether the \citet{snellen-etal-2010}
measurements could be explained by a greater abundance of CO on the
eastern terminator, where temperatures are warm and wind
preferentially flows from day to night, and a reduced abundance of CO
(and enhancement of CH$_4$) on the western terminator, where
temperatures are generally cooler.  This is unlikely, however, because
the timescales for chemical interconversion between CO and CH$_4$ in
the observable atmosphere are orders of magnitude longer than
dynamical timescales, so CO and CH$_4$ should be chemically quenched
\citep{cooper-showman-2006}.  Therefore, the abundance of CO should be 
essentially the same everywhere along the terminator at pressures
low enough to be sensed remotely.

Finally, while we have emphasized the wind patterns and implications
for transit Doppler measurements, the dynamics described here also 
predict a regime transition in the temperature structure that may
be important in explaining thermal observations from light curves
and secondary eclipses.  The shallow-water models in 
Figures~\ref{stswm-no-drag} and \ref{stswm-with-drag}, and
the 3D models in Figure~\ref{temp-winds}, show that the flow tends 
to a state with small
longitudinal temperature variations when radiation and friction are
weak, whereas the day-night
temperature differences become large when either radiation or
friction become strong.  Our models therefore predict a transition
from small to large fractional day-night temperature differences at
the infrared photosphere as stellar flux increases from small to large.  
We will explore this issue further in future work.


\acknowledgements
This research was supported by NASA Origins and Planetary Atmospheres
grants to APS.


\bibliographystyle{apj}
\bibliography{showman-bib}



\end{document}